\documentclass[11pt]{article}
\usepackage{natbib}
\usepackage[utf8]{inputenc}
\usepackage[T1]{fontenc}
\usepackage{mathptmx} 
\usepackage[margin=1in]{geometry}
\usepackage{amsmath}
\usepackage{amssymb}
\usepackage{mathtools}
\usepackage{amsthm}
\usepackage{amsfonts}
\usepackage{bm}
\usepackage{graphicx}
\usepackage{url}
\usepackage{hyperref}
\usepackage{cleveref}
\usepackage{xurl}
\usepackage{latexsym}
\usepackage{subcaption}
\usepackage{booktabs}
\usepackage{multirow}
\usepackage{multicol}
\usepackage{array}
\usepackage{caption}
\usepackage{rotating}
\usepackage{xcolor}
\usepackage{algorithm}
\usepackage{algpseudocode}
\usepackage{setspace}

\crefname{figure}{Figure}{Figures}

\hypersetup{
    colorlinks=true,
    linkcolor=blue,
    filecolor=black,
    citecolor=blue,
    urlcolor=black,
    pdftitle={Watermark in the Classroom: A Conformal Framework for Adaptive AI Usage Detection},
    pdfsubject={Large Language Models, Watermarking, AI in Education}
}

\title{Watermark in the Classroom: A Conformal Framework for Adaptive AI Usage Detection}
\author{
Yangxinyu Xie, Xuyang Chen, Zhimei Ren, Weijie~J. Su\thanks{University of Pennsylvania}
} 
\date{} 

\begin{document}

\maketitle

\begin{abstract}
As artificial intelligence tools become ubiquitous in education, maintaining academic integrity while accommodating pedagogically beneficial AI assistance presents unprecedented challenges. Current AI detection systems fail to control false positive rates (FPR) and suffer from bias against minority student groups, prompting institutional suspensions of these technologies. Watermarking techniques offer statistical rigor through precise $p$-values but remain untested in educational contexts where students may use varying levels of permitted AI edits.
We present the first adaptation of watermarking-based detection methods for classroom settings, introducing conformal methods that effectively control FPR across diverse classroom settings. Using essays from native and non-native English speakers, we simulate seven levels of AI editing interventions—from grammar correction to content expansion—across multiple language models and watermarking schemes, and evaluate our proposal under these different setups. Our findings provide educators with quantitative frameworks to enforce academic integrity standards while embracing AI integration in the classroom.
\end{abstract}
\vspace{0.2in}
\noindent\textbf{Keywords:} AI in Education, Academic Integrity, Large Language Model Watermarking, Conformal Prediction

\section{Introduction}

\subsection{Background and Motivation}

In August 2023, Vanderbilt University disabled Turnitin's AI detector \citep{TurnitinAIDetection} indefinitely, despite the growing pressure on institutions to maintain academic integrity in the age of generative artificial intelligence (AI) \citep{Coley2023AIDetection}. Soon after, universities across the country followed suit \citep{moorhouse2023generative, mcdonald2025generative, PleaseduAIDetectors}. 

The culprit was a loss of confidence in the tools' reliability. Commercial automated AI-text detectors---often built on training-based classifiers, such as Turnitin’s AI detector and GPTZero \citep{GPTZeroWebsite}---lack statistical guarantees regarding false-positive rates \citep{khalil2023will, perkins2024genai, Fowler2023ChatGPTDetector, Bahar2023AIDetectionFalseCase}. Furthermore, these tools have been shown to bias against non-native English speakers (NNES) and first-generation students \citep{liang2023gpt, campino2024unleashing, stone2024generative}. Students subjected to false accusations experience profound psychological distress, heightened anxiety, and diminished trust in educational institutions \citep{stone2023student, stone2024generative, Coldwell2024AICheating}. Moreover, the disproportionate impact on NNES and first-generation students can exacerbate existing educational inequities and may impede their academic progression \citep{wee2023non, stone2024generative, morris2019case}. 

\begin{figure}[thb!]
  \centering
  \includegraphics[width=0.5\linewidth]{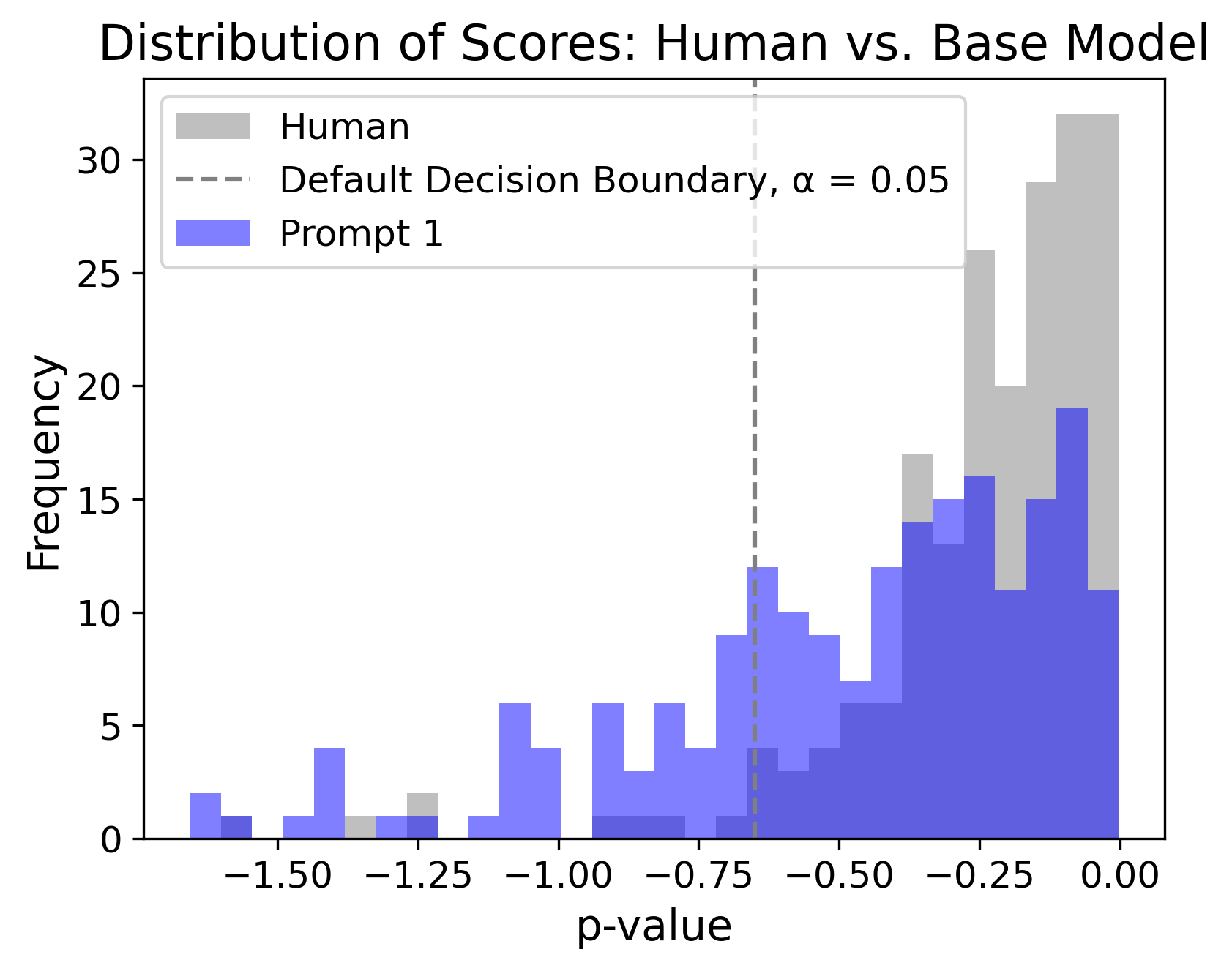}
  \caption{Distribution of watermarking $p$-values (on $\log_{100}$ scale) for human-written essays (gray) and essays with permitted AI grammar assistance (blue). Prompt 1 (blue) refers to prompting an AI model for minimal grammar correction; see definition in Section~\ref{sec:ai_assistance}.}
  \label{fig:pval_dist_grammar}
\end{figure}

\begin{figure}[thb!]
  \centering
  \begin{subfigure}[b]{0.48\linewidth}
    \includegraphics[width=\linewidth]{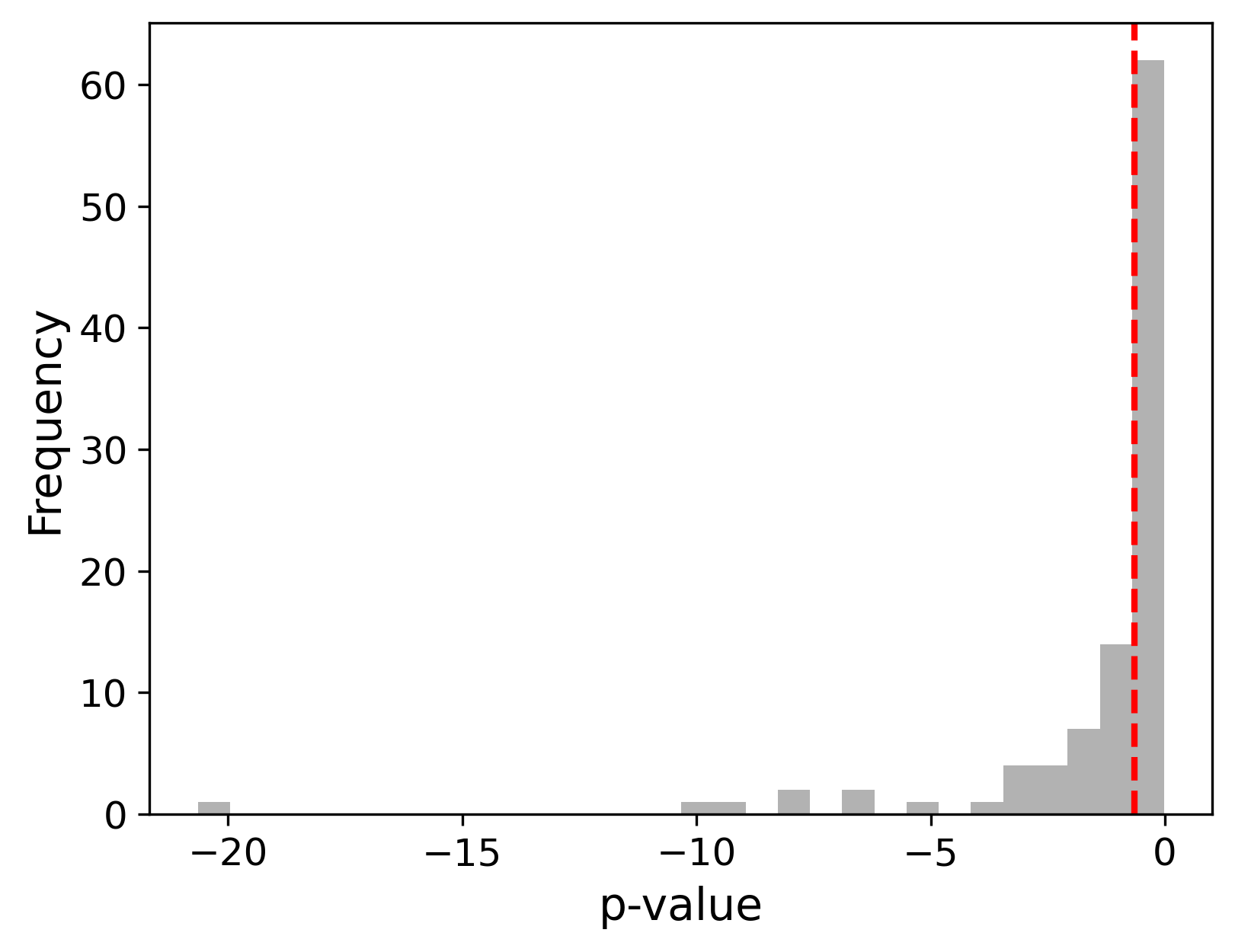}
    \caption{Initial view without labels}
    \label{fig:visual_inspection_a}
  \end{subfigure}
  \hfill
  \begin{subfigure}[b]{0.48\linewidth}
    \includegraphics[width=\linewidth]{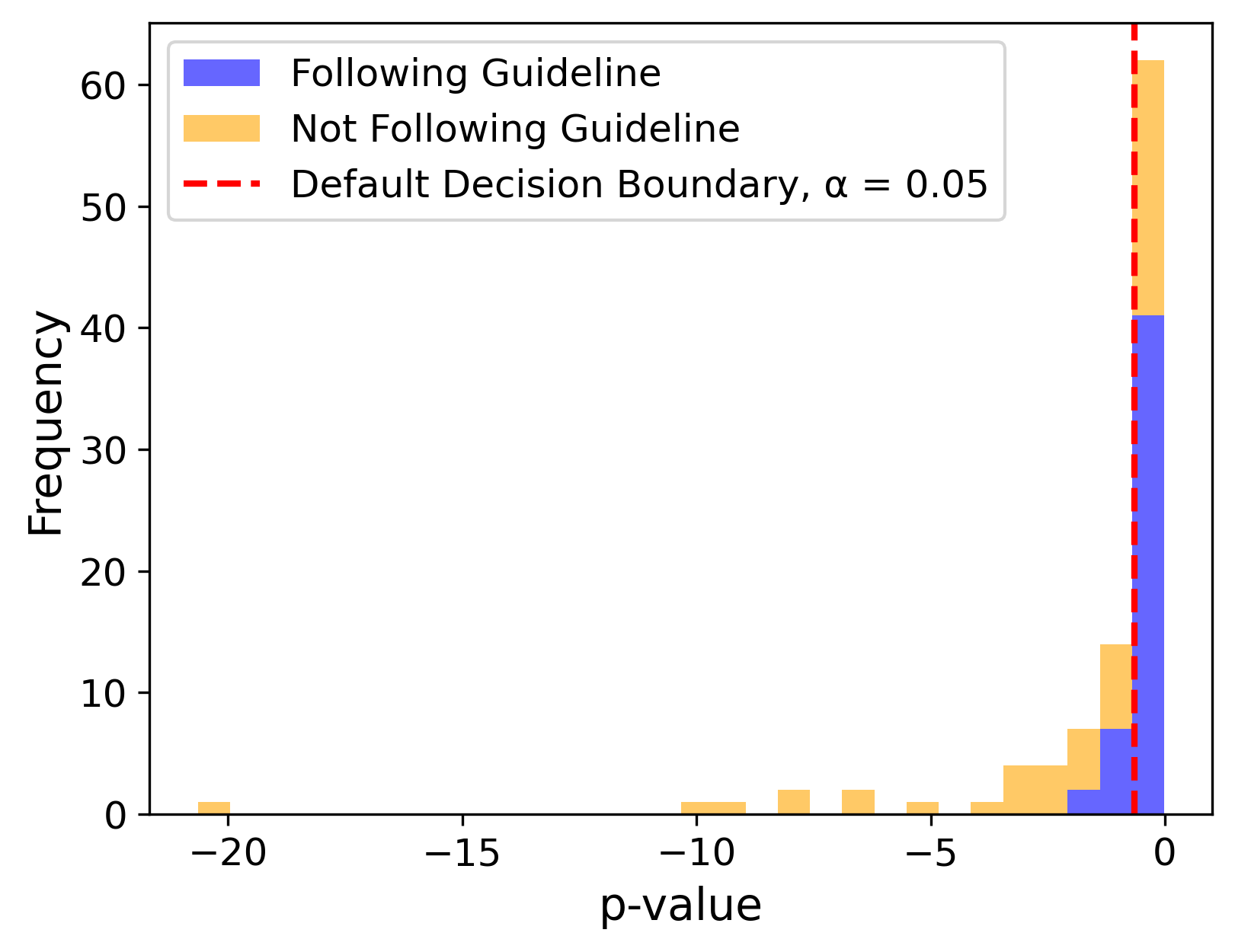}
    \caption{View with true labels}
    \label{fig:visual_inspection_b}
  \end{subfigure}
  \caption{Histogram of watermarking $p$-values (on $\log_{100}$ scale) across 100 simulated student submissions. We encourage the reader to examine panel (a) and determine a threshold that would optimally differentiate between permitted and prohibited AI usage. The right panel (b) reveals ground truth labels, demonstrating the inherent difficulty of threshold selection without proper calibration data. }
  \label{fig:visual_inspection}
\end{figure}

Meanwhile, generative watermarking techniques have emerged as a potentially promising alternative \citep{zhao2024sok, lancaster2023artificial, xie2023ai, majovsky2024perfect}. Unlike training-based classifiers, a watermarking protocol embeds verifiable statistical signals into model outputs during the generation process itself \citep{kirchenbauer2023watermark, aaronson, xie2024debiasing,abdalla2025llm}. This approach yields precise $p$-values against the null hypothesis that a piece of text is entirely human-written, offering a more rigorous statistical foundation for detection than the heuristic ``AI-likelihood'' scores that classifier-based solutions provide. 

Despite this theoretical promise, watermark detectors remain untested in classrooms \citep{perkins2024genai}. The challenge lies in the evolving nature of AI use in classrooms. Like classifier-based solutions, existing watermark techniques focus on distinguishing purely human-written texts versus machine-generated ones. However, as schools progressively integrate AI technologies into their pedagogy and encourage limited AI assistance to enhance student learning outcomes \citep{milano2023large, perkins2024artificial, cai2025exploring, Heaven2023ChatGPT, kangwa2025balancing, amato2025research, mollick2024instructors, nie2024gpt, eaton2025global}, educators need to distinguish between allowed and prohibited AI use—a nuanced task that current watermarking techniques cannot directly address. For example, if a professor allows AI-based grammatical editing but forbids content generation, a na\"{i}ve application—flagging any submission with watermark $p$-value below a fixed $\alpha$ threshold—can fail to maintain the expected false positive rate at $\alpha$. Figure~\ref{fig:pval_dist_grammar} illustrates this challenge through a simulation: while only 4\% of the human-written essays attain watermark $p$-values below the $\alpha = 0.05$ threshold, 25.5\% of the essays with permitted AI-grammar editing attain watermark $p$-values below the same threshold. In other words, a student who follows the AI use guidelines can easily be falsely accused without mechanisms to adjust the threshold.

Establishing valid thresholds that effectively control false positives is challenging. In large classrooms, instructors may collect many watermark $p$-values from student submissions, potentially enabling visual inspection of the histograms (Figure~\ref{fig:visual_inspection}) to make decisions. However, in the absence of prior knowledge, determining the cutoffs can be subjective and inconsistent. Moreover, such a problem can be further exacerbated in small classrooms, where the limited sample size renders visual inspection methods even more unreliable.

\subsection{This Article: A Conformal Framework for Adaptive AI Usage Detection}

Our analysis proceeds from the premise that hybrid writing content created through human-AI collaboration will become the new norm \citep{eaton2023postplagiarism, eaton2025global, luo2024critical}. As such, while existing commercial AI detection tools and research focus on the binary classification of solely human versus solely machine-generated content \citep{jiang2024detecting,hyatt2025using,TurnitinAIDetection, GPTZeroWebsite}, we examine the nuanced spectrum of human-AI collaboration.

Inspired by emerging pedagogical frameworks \citep{ogunleyea2024journal, perkins2024artificial, furze2024ai, kilincc2024comprehensive}, we consider the following setting: First, instructors provide explicit guidelines regarding permitted AI use for writing assignments. For example, instructors may outline a set of approved prompts to use when interacting with large language models (LLMs) to edit their essays. After students submit their assignments, instructors receive the corresponding scores from AI watermark detection tools. To ensure academic integrity, instructors are interested in which submissions may have involved more AI-edits than the guidelines allow. Nonetheless, comprehensively enumerating all possible guideline violations is intractable. Given this complexity, we propose the philosophy that submissions with AI-edit signals statistically indistinguishable from those produced under permitted guidelines should not trigger academic integrity violations. This approach prioritizes controlling the false positive rate (FPR) of detection methods, defined as the probability of incorrectly rejecting the null hypothesis that a submission adheres to AI usage guidelines, in parallel with recent research trends in education literature \citep{hyatt2025using, giray2024problem}.

A priori, the implementation of such a framework requires knowing the distribution of watermark scores under permitted AI usage patterns. In this work, we propose integrating conformal prediction methods, as they don't require explicit assumptions on the underlying distribution of the watermark scores—by using the historical data as a calibration set, we learn what ``normal'' looks like \citep{vovk2005algorithmic, angelopoulos2023conformal}.
However, the availability and composition of such calibration datasets can vary substantially across classrooms. The heterogeneous class sizes are one such variable: empirical data from representative American higher education institutions indicate that approximately 25\% of regular courses enroll fewer than 10 students, 27\% accommodate more than 50 students, and 47\% fall within the intermediate range of 10-40 students \citep{CollegeDataSize, WhartonFAQ}. This heterogeneity is further compounded by the diversity of student populations, which can include non-native English speakers, students with disabilities, and those from various academic backgrounds.

In Section \ref{sec:scenarios}, we discuss conformal methodologies to deliver rigorous, data-driven control of false-positive rates under three distinct classroom scenarios: large classrooms with similar past assignments, smaller classrooms with other sources of historical data, and adapting to heterogeneous student populations. In Section \ref{sec:methods}, we describe the methods to simulate AI-usage and detection in classrooms and present the empirical results in Section \ref{sec:results}. Limitations and future extensions are included in Section \ref{sec:discussion}.

\section{The Post-AI Classrooms: Scenarios and Detection Approaches}
\label{sec:scenarios}

In this section, we discuss three distinct scenarios in universities with corresponding conformal methodologies designed to rigorously control the FPR. These methods offer statistically robust solutions to the challenge of upholding academic integrity while embracing the pedagogical potential of regulated AI integration. We summarize the three scenarios and their corresponding conformal methods in Table~\ref{tab:scenarios} for ease of reference. 

\begin{table*}[t]
  \centering
  \begin{tabular}{p{0.4\linewidth}p{0.28\linewidth}p{0.24\linewidth}}
    \toprule
    Scenario & Data Requirement & Conformal Method \\
    \midrule
    Large class, similar past assignments & $n$ past essays, $n$ relatively large & Standard conformal prediction\\
    \hline
    Small classes, diverse past assignments & $K$ groups, $\{n_k\}_{k=1}^K$ essays per group & Hierarchical conformal prediction \\
    \hline
    Distribution shift, small subgroup & past essays from minority and majority student groups & Weighted conformal prediction \\
    \bottomrule
  \end{tabular}
  \caption{Comparison of conformal detection methods by classroom setting.}
  \label{tab:scenarios}
\end{table*}

\subsection{Large Classrooms with Similar Past Assignments}
\label{sec:scenario1}

If a professor teaches a large-enrollment class (e.g., 50+ students) and reuses essay writing prompts from previous semesters, it might be feasible to collect many past essays from students who adhered to the guidelines. For example, the professor may have access to $n$ essays from past cohorts prior to the popularization of accessible AI tools like ChatGPT \citep{achiam2023gpt}. In this case, the professor can simulate permitted AI usage by applying AI-edits to these essays per the guidelines. If the course has been rolled out for several years post AI, the professor may have access to many essays that were written by students who followed the guidelines, via record-tracking, auditing tools or honor code affirmation.

In this paper, we use $S = \{s_1,\dots,s_n\}$ to refer to the original $p$-values returned by watermarking detection providers, indicating the strength of evidence against the null hypothesis that a piece of text is entirely human-written. A lower watermark score indicates a stronger watermarking signal, suggesting more extensive AI edits.\footnote{Theoretical studies has shown that the watermark signal strength is a byproduct of the entropy of the sampling distribution during text generation process~\citep{kirchenbauer2023watermark,zhao2024sok,li2025statistical,abdalla2025llm}. In other words, the larger the set of plausible next words to generate from, the higher the watermark strength. When it comes to our setting, a larger set of possible edits leads to higher entropy during text generation, hence stronger watermark strength.} Given a new submission, the professor receives the watermark score $s$ and determines whether the essay is likely to be a violation of the guidelines by computing the conformal value $p$ \citep{vovk2003testing,vovk2005algorithmic}:

\begin{equation}
\hat u(s) = \frac{1 + \sum_{i=1}^n \mathbf{1}\{s_i \le s\}}{n+1}.
\end{equation}

\begin{figure}
    \centering
    \includegraphics[width=\textwidth]{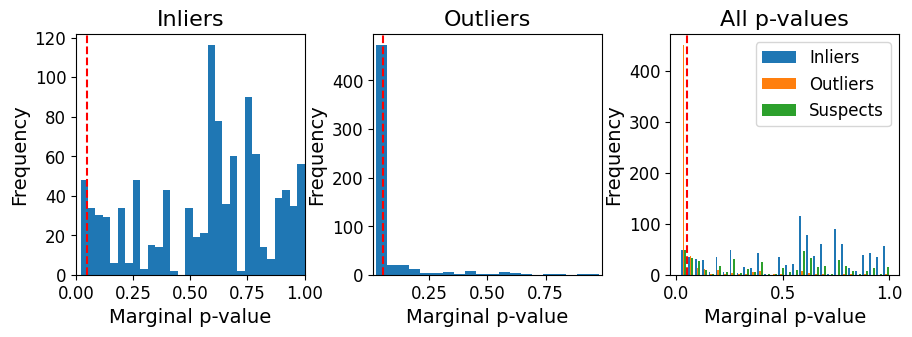}
    \caption{Distribution of conformal $p$-values. \textbf{Left:} Essays following guidelines (Inliers) exhibit approximately uniform distribution of $p$-values between 0 and 1, with only 4.69\% falling below $\alpha = 0.05$ (red line)—confirming the theoretical guarantee on false positive rate. \textbf{Middle:} Essays with clear violations (Outliers) show a sharp concentration of $p$-values near zero, indicating high detectability. A clear violation describes an essay that violates the guidelines and undergoes substantial AI-induced changes. \textbf{Right:} Combined view revealing how essays following guidelines (blue), clear violations (orange), and ``suspect'' cases (green) occupy distinct regions in the conformal $p$-value space, enabling robust classification with statistical guarantees. Briefly speaking, ``suspect'' refers to essays which also do not follow guidelines but involve only minor AI-edits. The formal definition of ``Outliers'' and ``Suspects'' could be found in Appendix~\ref{sec:refine_def}.} 
    \label{fig:conformal_pvals}
\end{figure}

This conformal $p$-value has a fundamentally different interpretation from the watermark $p$-value. It represents: ``What proportion of essays with permitted AI edits have watermark signals at least as strong as this submission?'' This way, the professor may flag $X$ as a likely violation if $\hat u(s) \le \alpha$, where $\alpha$ is the target false positive level (e.g., 0.05). This method guarantees that, for independent, identically distributed (i.i.d.) essays genuinely following permitted AI usage, $\Pr(\hat u(s) \le \alpha) \le \alpha,$ regardless of the underlying distribution of watermark scores. In other words, if the new essay is indeed written by a student who follows the guidelines, the probability of the watermark score being below the threshold $\alpha$ is at most $\alpha$. An example of this is shown in Figure~\ref{fig:conformal_pvals}.

\subsection{Small Classrooms with Limited I.I.D. Data}

\label{sec:scenario2}

While the previous method provides a statistically rigorous approach, it requires a fairly large number of calibration essays similar enough to the new submission to be effective. Suppose a professor teaches a small seminar course with only 10-20 students, limiting her access to a large set of i.i.d. calibration data from past cohorts, if any. Moreover, she may update the assignments to reflect the latest developments in the field, or teach a new course with no prior data. 

Nevertheless, many past assignments sharing common educational objectives or quantitative skills may be readily accessible to instructors, and these naturally form groups of related but non-identically distributed assignments. Hierarchical conformal prediction extends the standard conformal framework to effectively handle such grouped data \citep{lee2023distribution}, assuming hierarchical exchangeability. This assumption has two components: (1) the groups themselves must be exchangeable with one another, meaning that we could randomly reorder the groups without affecting our analysis; and (2) the individual observations within each group must also be exchangeable, meaning we could randomly reorder essays within any single group. In other words, to meet this assumption, though each assignment can cover different writing topics, it is intended for similar learning prerequisites and outcomes (assumption (1)), and given to students who completed the assignments individually (assumption (2)).

\begin{figure}
    \centering
    \includegraphics[width=\textwidth]{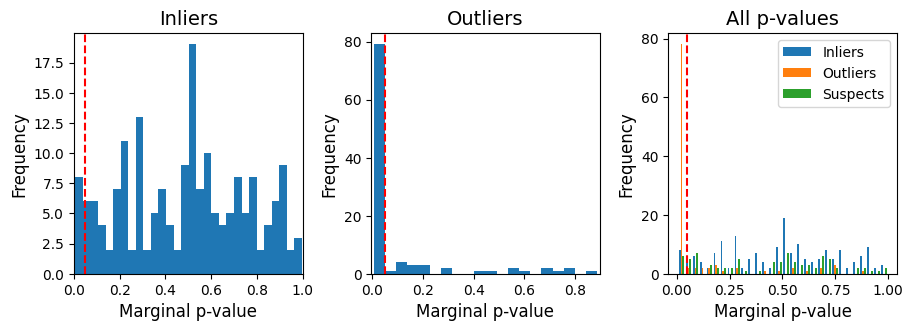}
    \caption{Distribution of hierarchical conformal $p$-values. \textbf{Left:} Essays following guidelines show $p$-values  with only 4.32\% falling below. \textbf{Middle:} Essays with clear violations show extreme concentration of $p$-values below 0.05, with a different x-axis scale highlighting the separation. \textbf{Right:} Combined view showing inliers (blue), outliers (orange), and suspects (green) across the full $p$-value range, demonstrating that hierarchical conformal prediction maintains discriminatory power even with diverse prompts. The red dashed line indicates the $\alpha = 0.05$ threshold.}
    \label{fig:hierarchical_pvals}
\end{figure}

Suppose that the professor has $K$ groups of calibration essays, each of size $n_k\, (k = 1,...,K)$. The watermark scores of these groups of essays are denoted as $S_k = \{s_{k,1},\dots,s_{k,n_k}\}$. The hierarchical conformal $p$-value for a new essay and its corresponding watermark score $s$ (not belonging to any aforementioned groups) is defined as:
\begin{equation}
\hat u_{\text{hier}}(s) = \frac{1}{(K+1)}\left[1 + \sum_{k=1}^{K}\frac{\sum_{i=1}^{n_k} \mathbf{1}\{s_{k,i} \leq s\}}{n_k}\right].
\end{equation}
This hierarchical conformal $p$-value places different amounts of weight on different calibration data points depending on the sizes of the various groups. This probability accounts for randomness at two levels: how the assignment is designed, and how the students completed this assignment. If the professor designs a new assignment designed with similar educational goals as the past classes included in the calibration set, if new student follows the AI guideline, the probability of false positive rate satisfies $\Pr(u_{\text{hier}}(s) \le \alpha) \le \alpha$.
An illustration of this approach is shown in Figure~\ref{fig:hierarchical_pvals}.

\subsection{Fairer Classrooms: Distribution Shift with Small Subgroups}
\label{sec:scenario3}

\begin{figure}
    \centering
    \begin{subfigure}[b]{0.48\textwidth}
        \centering
        \includegraphics[width=\textwidth]{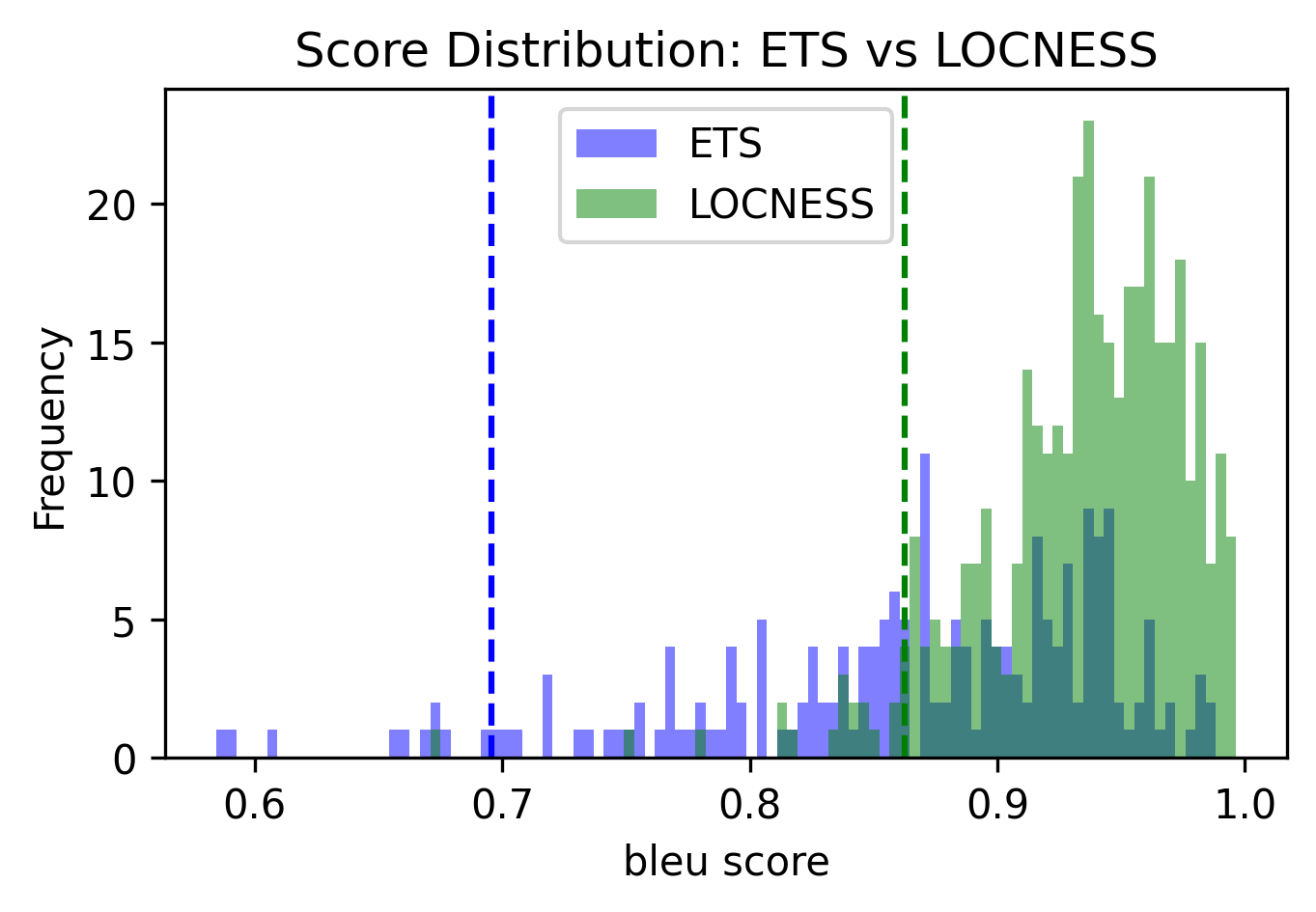}
        \caption{BLEU score distributions}
        \label{fig:dist_shift_a}
    \end{subfigure}
    \hfill
    \begin{subfigure}[b]{0.48\textwidth}
        \centering
        \includegraphics[width=\textwidth]{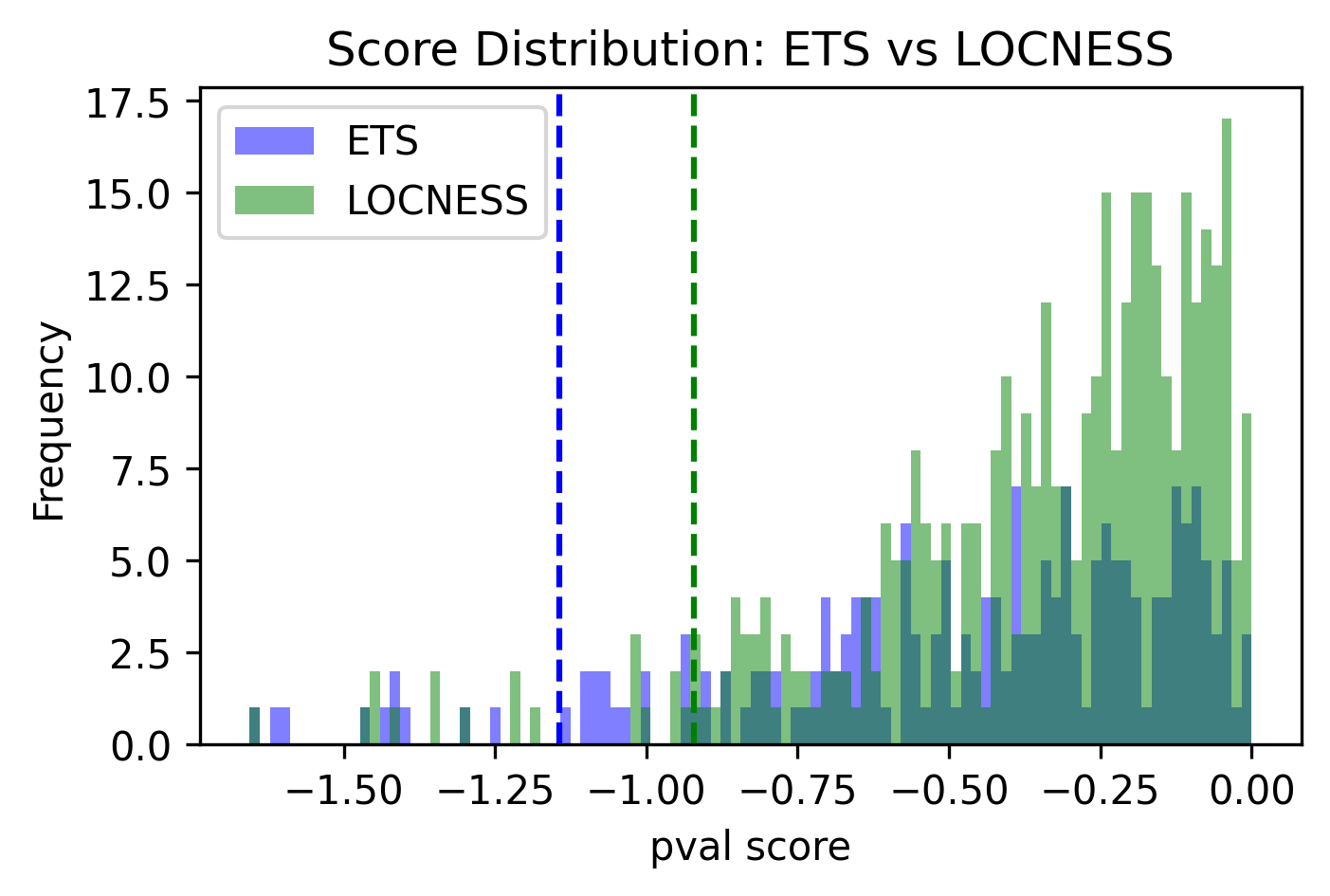}
        \caption{Watermarking $p$-value distributions}
        \label{fig:dist_shift_b}
    \end{subfigure}
    \caption{Comparative analysis of AI assistance impact across language backgrounds. (a) BLEU score distributions showing the similarity between original and AI-edited essays for ETS corpus (non-native speakers, blue) and LOCNESS corpus (native speakers, green). Non-native speakers' essays show substantially more AI modifications with BLEU scores extending to the left of the blue dashed line. (b) Corresponding watermarking $p$-value distributions ($\log_{10}$ scale) showing stronger watermarking signals (lower $p$-values) for non-native speakers compared to native speakers when using identical AI assistance. The blue and green vertical dashed lines indicate recommended decision boundaries for each population.}
    \label{fig:dist_shift}
\end{figure}

In many cases, the calibration data may not be i.i.d. due to distribution shifts, such as when a professor teaches a course with a diverse student population, including non-native English speakers (NNES) or students with disabilities. Previously, studies reported that NNES students are more likely to be flagged as AI-generated by existing AI detectors \citep{wee2023non, liang2023gpt, campino2024unleashing}. Introducing fair detection methods has become a pressing need in the field of education \citep{jiang2024towards, jiang2024detecting}.

When using AI for permitted grammatical assistance, essays by non-native speakers can show stronger watermarking signals compared to those by native speakers. As shown in Figure~\ref{fig:dist_shift}, with the same essay-editing prompt, the watermarking $p$-values for NNES students (blue) are significantly lower than those for native speakers (green), due to the higher degree of AI modifications in their essays. If the professor uses the same detection threshold for both groups, the FPR for NNES students will be much higher than that for native speakers, leading to unfair treatment. 

In addition, even though the professor has access to calibration data from both groups, the number of essays from NNES students can be much smaller than that from native speakers. In this case, only using the calibration data from NNES students may not be sufficient to provide a reliable threshold due to the small sample size. To address this issue, we propose a weighted conformal prediction method that combines the calibration data from both groups while accounting for the distribution shift \citep{tibshirani2019conformal}.

Let $S = \{s_1,\dots,s_n\}$ be the watermark scores of the calibration essays and $s$ that of a new submission. We introduce the notion of normalized importance weights as follows:
$$
w_i(s) = \frac{\hat q(s_i)/\hat p(s_i)}{\sum_{j=1}^n \hat q(s_j)/\hat p(s_j) +\hat q(s)/\hat p(s) }, \quad w(s) = \frac{\hat q(s)/\hat p(s)}{\sum_{j=1}^n \hat q(s_j)/\hat p(s_j) +\hat q(s)/\hat p(s) },
$$
where $\hat p$ and $\hat q$ are the density estimates for the full and target-group distributions, respectively. 
In other words, $p$ is the distribution of all the calibration essays, and $q$ is the distribution of the smaller target group (e.g., NNES students). For a new essay from a non-native English speaker, we compute a weighted $p$-value by comparing its watermark score $s$ to the scores of the calibration essays, weighted by $w_i$. We flag the new essay as a likely violation if the weighted sum of the calibration essays with conformity scores less than or equal to $s$ is below a chosen significance level $\alpha$, i.e.:
$$
\sum_{i=1}^n w_i(s) \cdot {\bf 1}\{ s_i \leq s \} + w(s) < \alpha.
$$

Under the assumption of weighted exchangeability \citep{tibshirani2019conformal, angelopoulos2024theoretical}, this method retains the guarantee as before. In this paper, since we assume minimal available information (i.e. only the scores $s$), we adapt kernel density estimation for $\hat p$ and $\hat q$ respectively, with details given in Section \ref{sec:experiment_setups} and  Appendix \ref{sec:quantile_shift}. When higher dimensional information is available, machine-learning-based approaches can be adopted to reach a more accurate estimate of the likelihood ratio $\hat r = \widehat{q(s_i)/p(s_i)}$ and avoid the possible instability of the kernel density estimation approach \cite{tibshirani2019conformal}. Furthermore, even with imperfect estimates of this likelihood ratio, the validity of the weighted conformal methods can still be guaranteed, with the coverage error rate controlled by the error of the estimate; we refer the reader to \cite{lei2021conformal} for more details.

In this paper, we present our framework through single-hypothesis testing—detecting AI-usage violations by individual students—to establish the fundamental principles. When examining potential violations among a group of students, our approach can be extended against multiple hypotheses straightforwardly to further control the false positive rate \citep{dunn1961multiple, benjamini1995controlling, ren2021model}.
\section{Simulating Classrooms}
\label{sec:methods}
This section summarizes the datasets, watermarking schemes, AI‐assistance interventions, calibration/test splits, and evaluation metrics used in our experiments. Detailed descriptions and additional results are provided in Appendix~\ref{sec:methods}.

\subsection{Datasets, Language Models and Watermarking Schemes}
We simulate classroom scenarios outlined in Section~\ref{sec:scenarios} using two publicly available corpora. After filtering and preprocessing, we use the following datasets: (1) \textbf{ETS TOEFL11 (ETS) Corpus} \citep{blanchard2013toefl11}, consisting of 8,884 essays by non-native speakers across eight writing prompts; (2) \textbf{LOCNESS Corpus} \citep{granger1998computer}, including 385 essays by native speakers covering 273 unique topics. 
For each essay in both datasets, we generate AI-edited versions by prompting two large language models (LLMs): \texttt{Phi-4-mini-instruct} \citep{abdin2024phi} and \texttt{Qwen2.5-7B-Instruct} \citep{yang2024qwen2}, applying two watermarking methods: Gumbel-Max \citep{aaronson} and Green/Red List \citep{kirchenbauer2023watermark}. In the main text, we primarily present the results using {Phi-4-mini-instruct} model with Gumbel-Max watermarking method. The results using other models and watermarking methods present similar patterns and are included in Appendix~\ref{sec:dataset_LM_watermark}.

\subsection{AI‐Assistance Interventions}
\label{sec:ai_assistance}
The following seven prompt templates cover a range of AI-edit interventions, from basic grammar checking to more complex tasks such as improving clarity and logical flow. In our simulation study, each prompt, coupled with the original essay, is fed into the LLMs to produce an AI-edited version. All prompts except for the last represent forms of AI assistance that could plausibly be allowed in the classroom. The last prompt is the strongest form of AI-usage guideline violation in our study, which students may use to generate an essay from scratch.

\begin{enumerate}
    \item ``Correct only spelling and grammatical errors in the following essay. Do not change wording, phrasing, punctuation (unless part of a grammar fix) or style.''
    
    \item ``Correct grammar, spelling, and any awkward sentence constructions in the following essay. Keep the original voice and tone. Do not rewrite for clarity unless absolutely necessary.''
    
    \item ``Improve clarity in the following essay by reordering or rephrasing within paragraphs only. Do not cut or add content. Keep the original sentences where possible.''
    
    \item ``Assess the logical flow of this essay and revise it to improve coherence or address reasoning gaps. Keep the original wording and structure as intact as possible.''
    
    \item ``Revise the following essay to enhance readability and clarity, while keeping the original wording as intact as possible.''
    
    \item ``Please revise the following essay to improve sentence transitions and flow. Do not add new content or examples—only improve how existing ideas are connected and expressed.''
    
    \item ``Expand on the essay below to complete it.'' For writing-prompt-specific essays (ETS corpus), we use: ``Help me write an essay responding to the following prompt: [prompt]. Here is the text I have written so far as a starting point: [essay]. Please build upon my ideas, keeping the original tone, style, and viewpoint.''
\end{enumerate}
The instructor-permitted prompt is defined as the ``null'' edit. For example, if prompt 1 is permitted and thus used for calibration, essays edited with prompt 1 are considered as the ``null'' edit and those with prompts 2–7 are ``alternative'' edits.

\subsection{Experiment Setups}
\label{sec:experiment_setups}
We evaluate our three conformal prediction methods (standard, hierarchical, weighted) under the classroom scenarios outlined in Section~\ref{sec:scenarios}, at a target significance level of $\alpha=0.05$. Each experiment is repeated with five random seeds in sampling the calibration set. 

\begin{enumerate}
\item Scenario 1 (Standard Conformal).
For each of the eight writing prompts in the ETS corpus, we:
\begin{itemize}
  \item Randomly hold out 200 essays, from which we sample $n_{\text{cal}}\in\{30, 50, 200\}$ essays for calibration.\footnote{The numbers are chosen to reflect the different class sizes discussed in the Introduction.} For each essay in the calibration set, we generate AI-edited versions using one of prompts 1–6 (see Section~\ref{sec:ai_assistance}) and compute the corresponding watermarking $p$-values.
  \item Use the remaining essays from the same writing prompt for testing. Apply prompts other than the calibration prompt and compute the corresponding watermarking $p$-values (e.g., if prompt 1 is used for calibration, test on essays edited with prompts 2–7). Table~\ref{tab:ets_test_size} reports test set sizes.
\end{itemize}

\begin{table}[htb!]
\centering
\begin{tabular}{|c|c|c|c|c|c|c|c|c|}
\hline
 & \textbf{P1} & \textbf{P2} & \textbf{P3} & \textbf{P4} & \textbf{P5} & \textbf{P6} & \textbf{P7} & \textbf{P8} \\ \hline
\textbf{Test Set Size} & 1024 & 962 & 783 & 887 & 997 & 492 & 1076 & 1063 \\ \hline
\end{tabular}
\caption{Test set sizes for each AI-assistance intervention in the ETS corpus. Each column corresponds to a different essay writing prompt (P1–P8) from the ETS corpus (this is not the AI-assistance prompts in Section~\ref{sec:ai_assistance}). }
\label{tab:ets_test_size}
\end{table}

\item Scenario 2 (Hierarchical Conformal).
We use the LOCNESS corpus for this scenario. We group the essays based on the 273 different topics, and apple hierarchical conformal methods to account for this structure. We randomly hold out 200 essays reserved for calibration and use the other 185 essays for testing. To consider different levels of historical data availabilities, we sample $n_{\text{cal}}\in\{30, 50, 200\}$ essays from the held-out dataset for calibration. All remaining steps follow Scenario 1.

\item Scenario 3 (Weighted Conformal).
We combine the ETS corpus (minority) and the LOCNESS corpus (majority):
\begin{itemize}
  \item Randomly sample $m\in\{5,15,30\}$ non‐native (ETS) essays and all $n=385$ native essays (LOCNESS) for calibration.
  \item Estimate densities $\hat p$ and $\hat q$. We estimate $\hat p$, the density of the distribution of the watermark scores of all calibration essays via Gaussian Kernel Density Estimation (KDE) \citep{silverman2018density} with a bandwidth of 0.5. Since the number of samples in the minority group is very small ($m\leq 30$), it is not feasible to capture the entire distribution shift without sufficient prior knowledge. In our simulation, we propose two methods:
  \begin{itemize}
    \item \textbf{Mean Shift}: The method calculates sample means for both the entire calibration set (the LOCNESS corpus and the $m$ sampled non-native essays) and the minority set (only the $m$ non-native essays). The shift is then computed as the difference between these two means. $\hat q$ is estimated by shifting $\hat p$ using this difference and the appropriate standard deviations of both sets.
    \item \textbf{Quantile Shift}: Instead of using the mean, this method computes a quantile that is as close to $\alpha$ as possible. The details are included in Appendix~\ref{sec:quantile_shift}.
  \end{itemize}
  \item Test on the same sets as in Scenario 1.
\end{itemize}
\end{enumerate}

\subsection{Evaluation Metrics}
Our evaluation focuses on two primary metrics across methods and scenarios: the false positive rate (FPR) and detection power (Power).

\begin{itemize}
  \item \textbf{False Positive Rate (FPR)}: The proportion of essays that are classified as outliers when they are following the permitted guidelines. This is computed as the number of flagged essays divided by the total number of essays in the test set while applying the same AI-edits as the calibration set. 
  
  \item Detection \textbf{Power}: The proportion of essays that are correctly classified as outliers.  We define `outliers' as essays resulting from an editing prompt not included in the guideline and exhibiting substantial AI-edits. On the other hand, if a resulting essay involves only minor edits, we refer to it as a `suspect' (see Appendix~\ref{sec:refine_def} for formal definitions).
\end{itemize}

\begin{figure}[htb!]
  \includegraphics[width=\textwidth]{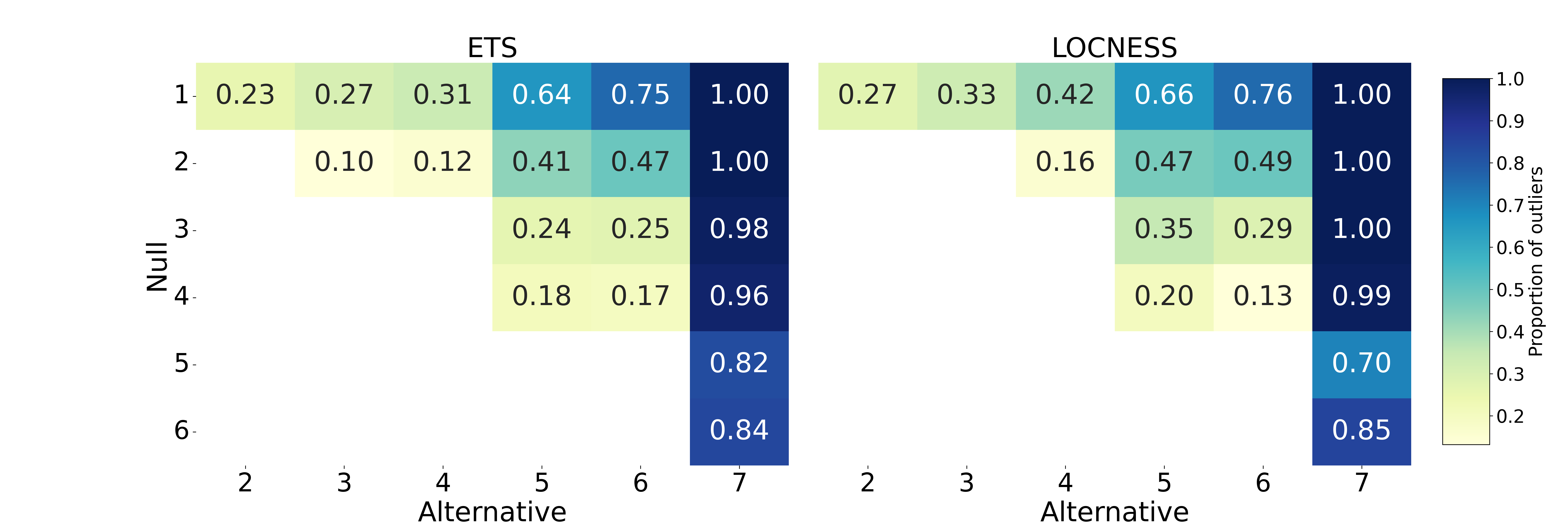}
  \caption{Average proportion of adjusted outliers when comparing each allowed ``null'' edit (rows, prompts 1–6) against ``alternative'' edits (columns, prompts 2–7). If the proportion is below 0.05, or its raw count is under 30, we consider it a negligible violation, and thus do not include it in the figures. These proportions are computed when AI-edits are made through the \texttt{Phi-4-mini-instruct} model and the Gumbel-Max watermarking method.}
  \label{fig:outlier_proportion}
\end{figure}

For scenario 1 and 3, we first compute the FPR and power for each of the eight writing prompts in the ETS corpus and average them across all writing prompts.\footnote{The same averaging is applied to compute the proportion of outliers in Figure~\ref{fig:outlier_proportion}.} Moreover, for all scenarios, each metric is averaged over the five repeated experiments with different random seeds. 

In our subsequent experiments, we operate under the assumption that most professors would be primarily concerned with identifying clear violations rather than ``suspect'' cases, as the latter could still demonstrate substantial student engagement with course material despite technical guideline infractions. We emphasize that instructors do not have access to students’ original drafts and thus cannot compute BLEU in practice. We use BLEU here solely to quantify and correct our reported FPR in simulations, ensuring that minor infractions do not unduly inflate the measured FPR.

Figure~\ref{fig:outlier_proportion} shows an example average proportion of adjusted outliers when comparing each allowed ``null'' edit (rows, prompts 1–6) against “alternative” edits (columns, prompts 2–7). We remark that some differences in the proportions of outliers in Figure~\ref{fig:outlier_proportion} are not statistically significant, as they are all below 0.05 or have raw counts below 30. These differences are left blank in the figure. In the following sections, we exclude these cases when reporting the results of our conformal prediction methods, to ensure accurate estimate of the false positive rates (FPR) and detection power. 

The empirical results reveal interesting relationships between AI-edit instructions and the resulting extent of AI involvement. In this example, the AI instruction prompt for improving clarity (prompt 3) and assessing logical flow (prompt 4) lead to substantially lower AI edits than those for improving readability, clarity (prompt 5), and flow (prompt 6). This suggests that mere structural logical improvements may involve less detectable AI modification than surface-level stylistic polishing. This, for example, may be a positive finding for educators who encourage AI assistance for brainstorming and critical thinking \citep{mollick2024instructors,perkins2024genai}.

Moreover, the detection boundaries reveal practical consequences of different permissible AI-assistance. For instance, permitting logical flow adjustments essentially renders grammar-level improvements undetectable as violations, as AI models tend to automatically correct grammar and spelling errors. Conversely, restricting permissions to strict grammatical correction preserves substantial possibility of correct detection against intermediate violations such as enhancing readability or transition. This trade-off between pedagogical flexibility and enforcement capability represents a fundamental constraint that educators must navigate when designing AI integration policies. In practice, the differing levels of AI-edits resulting from the different AI-assistance interventions may not be obvious a priori. Thus, real world deployment may benefit from similar detection boundary analyses to formulate an informed AI-assistance guideline.
\section{Results}
In this section, we present the experimental results of three conformal prediction methods (standard, hierarchical, and weighted) and analyze their false discovery rates and detection powers.
\label{sec:results}

\subsection{Scenario 1: Standard Conformal Method}
\label{sec:results_scenario2}

\begin{figure}
\centering
\includegraphics[width=0.95\textwidth]{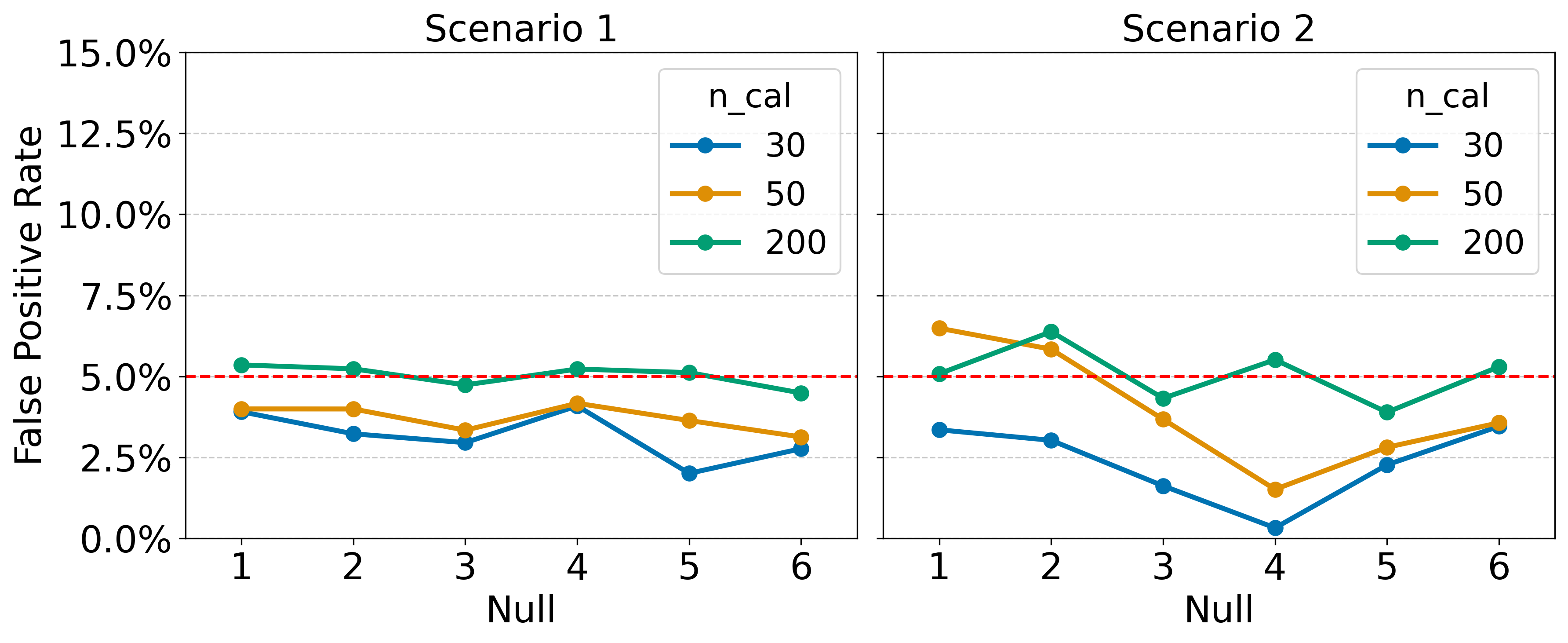}
\caption{Average FPR in Scenario 1 and Scenario 2.}
\label{fig:conformal_fpr}
\end{figure}

As shown in Figure~\ref{fig:conformal_fpr} (Left), the standard conformal method effectively controls the FPR across AI-usage guidelines, maintaining it around or below the 5\% significance level on average. When the number of calibration essays increases, the FPR shows a trend of converging towards the nominal 5\% level, indicating the benefit of larger calibration sizes, resource permitting.

\begin{figure}
    \centering
    \includegraphics[width=\textwidth]{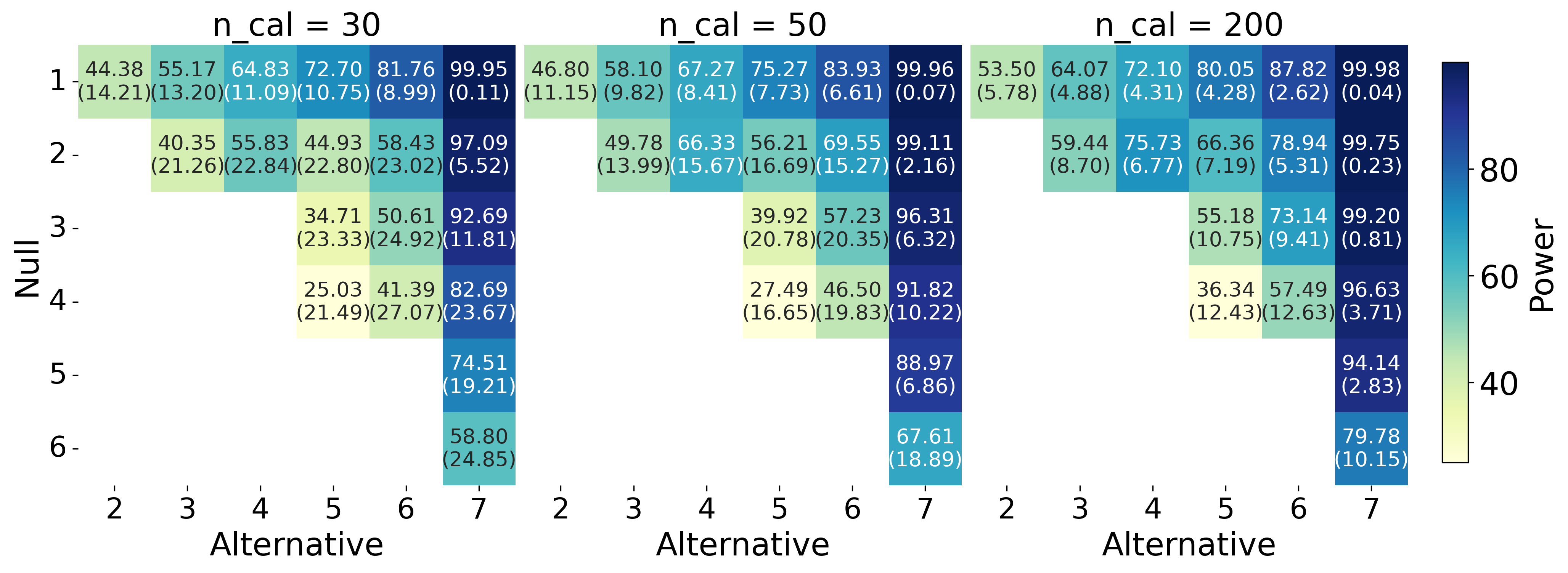}
    \\
    \includegraphics[width=\textwidth]{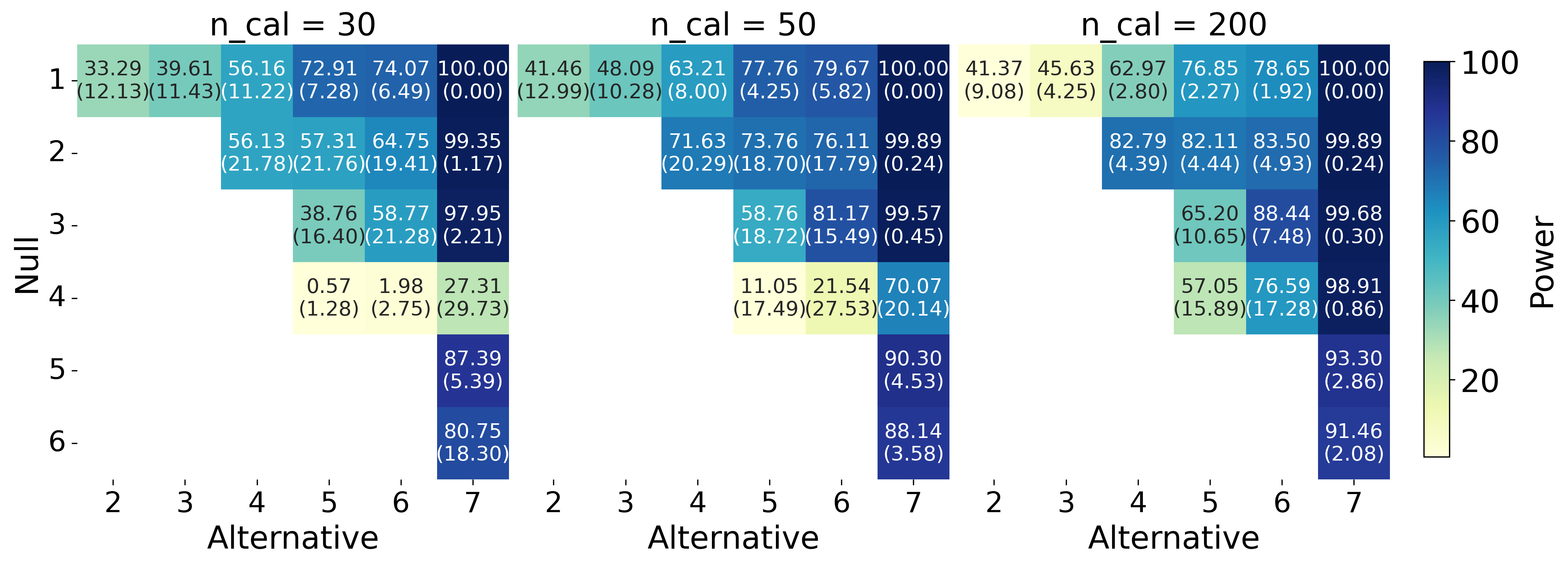}
    \caption{Detection Power. Top: Scenario 1 (ETS), Standard Conformal Method. Bottom: Scenario 2 (LOCNESS), Hierarchical Conformal Method.}
    \label{fig:conformal_power_ets}
\end{figure}

Figure~\ref{fig:conformal_power_ets} (Top) plots detection power of the standard conformal method by null and alternative edit levels across calibration set sizes $n_{\mathrm{cal}}=30,50,200$. The general trend of detection power parallels the proportion of outliers in the test sets (Figure \ref{fig:outlier_proportion}). In particular, if there is a strong violation of the AI usage guideline, such as extreme edits (null=1 vs.\ alt=7), the detection power is close to 100\% even at relatively modest calibration sizes (e.g., $n_{\mathrm{cal}}=30$). However, as the permitted AI involvement increases, the detection power (e.g., null=6 vs.\ alt=7) decreases significantly, especially at smaller calibration sizes. Moreover, more subtle guideline violations (e.g., null=1 vs.\ alt=2) are harder to detect, with power dropping below 50\% at $n_{\mathrm{cal}}=30$ and remaining low even at larger calibration sizes. The results suggest that while the conformal method is effective in detecting extreme violations, the instructor may need to balance the permitted AI involvement in the guidelines with the detection power. Meanwhile, since detection power increases as the calibration set size increases, the instructor may consider collecting more calibration data to improve detection power, especially for more subtle violations or more permitting AI involvement.

\subsection{Scenario 2: Hierarchical Conformal Method}
\label{sec:results_scenario3}
    Figure~\ref{fig:conformal_fpr} (Right) indicates the same pattern as the standard conformal method, demonstrating the effectiveness of the hierarchical conformal method in controlling the FPR.
    
    Figure~\ref{fig:conformal_power_ets} (Bottom) shows the detection power of the hierarchical conformal method across different calibration set sizes and AI-usage guidelines. The detection power generally follows the same trend as the standard conformal method, with a notable difference. The hierarchical method tends to have lower detection power than the standard method if the calibration set size is small (e.g., $n_{\mathrm{cal}}=30$). This is due to the hierarchical structure of the data, violating the exchangeability assumption of the standard conformal method. If more data are clustered together, the hierarchical method can still maintain valid FPR control, but at the cost of detection power. 
\subsection{Scenario 3: Weighted Conformal Prediction}
\label{sec:results_scenario4}

We compare the weighted conformal method with two unweighted methods, which do not account for the distribution shift between the calibration and test sets:
\begin{enumerate}
    \item \textbf{In Dist. Only}: The method only uses the calibration set from the in-distribution essays, ignoring the larger volume of native-speaker essays. When the calibration set size $m$ is small, this method is useless when $\alpha$ is less than one divided by the $1/m$ (e.g., $\alpha =0.05, m=5$ or $15$).
    \item \textbf{Comb. Unweighted}: The method combines the calibration sets from both native-speaker and non-native-speaker essays without weighting, treating them equally.
    \item \textbf{Comb. Weighted}: The method combines the calibration sets from both native-speaker and non-native-speaker essays, but applies a weighting scheme to account for the distribution shift. The two weighting schemes are discussed in Section~\ref{sec:experiment_setups}.
\end{enumerate}
Table~\ref{tab:conformal_results_phi_openai} presents the performance of three conformal prediction methods under varying NNES calibration sizes $m$ and permitted levels of AI involvement. Both weighted methods lowers the FPR compared to the Comb. Unweighted method, even though at times the FPR may not be as close to the nominal level of 5\% as the In Dist. Only method, especially when the calibration set size is larger (e.g., $m=30$). This is likely due to the imperfect estimate of $q$, and thus the density ratio between the calibration and test sets. With substantial prior knowledge and more student information, the instructor may develop a more accurate estimate of the density ratio, which can further improve the FPR control.

The weighted conformal method generally achieves lower detection power than the Comb. Unweighted method. This is expected as the weighted method is designed to account for the distribution shift between the calibration and test sets, which may lead to a more conservative estimate of the detection power. However, the weighted method still maintains competitive detection power across different calibration set sizes and AI-usage guidelines. The detection power generally follows the same trend as the unweighted methods, with lower power for more subtle violations of the AI usage guidelines or more permissive AI involvement. 
\begin{table}[h!]
\tiny
\centering
\caption{Weighted Conformal Methods: FPR and Power for Different Calibration Sizes and AI-Usage Guidelines. The FPR closest to $\alpha=0.05$ is highlighted in bold. }
\label{tab:conformal_results_phi_openai}
\resizebox{\textwidth}{!}{%
\begin{tabular}{c c c | c c | c c | c c | c c }
\toprule
Null & Alt. & $m$ & \multicolumn{2}{c|}{In Dist. Only} & \multicolumn{2}{c|}{Comb. Unweighted} & \multicolumn{4}{c}{Comb. Weighted}\\
& & & & & & & \multicolumn{2}{c|}{Quantile} & \multicolumn{2}{c}{Mean} \\
& & & FPR & Power & FPR & Power & FPR & Power & FPR & Power \\
\midrule
1 & 7 & 5 & 0.0 & 0.0 & 12.7 & 100.0 & \textbf{6.4} & 100.0 & 9.7 & 100.0 \\
~ & ~ & 15 & 0.0 & 0.0 & 12.7 & 100.0 & 8.4 & 100.0 & \textbf{7.9} & 100.0 \\
~ & ~ & 30 & 3.4 & 99.9 & 11.8 & 100.0 & \textbf{6.6} & 100.0 & 6.7 & 100.0 \\
\midrule
2 & 7 & 5 & 0.0 & 0.0 & 13.2 & 100.0 & \textbf{5.8} & 99.4 & 10.1 & 99.9 \\
~ & ~ & 15 & 0.0 & 0.0 & 13.0 & 100.0 & \textbf{6.9} & 99.5 & 7.9 & 99.6 \\
~ & ~ & 30 & \textbf{3.5} & 98.4 & 12.1 & 100.0 & 6.8 & 99.7 & 7.1 & 99.8 \\
\midrule
3 & 7 & 5 & 0.0 & 0.0 & 9.4 & 99.8 & \textbf{4.7} & 97.3 & 7.5 & 99.6 \\
~ & ~ & 15 & 0.0 & 0.0 & 9.6 & 99.8 & 6.8 & 98.5 & \textbf{6.5} & 97.8 \\
~ & ~ & 30 & 3.0 & 91.6 & 9.1 & 99.8 & \textbf{5.7} & 97.5 & 5.8 & 97.7 \\
\midrule
4 & 7 & 5 & 0.0 & 0.0 & 8.0 & 99.1 & \textbf{4.4} & 80.5 & 7.1 & 97.3 \\
~ & ~ & 15 & 0.0 & 0.0 & 8.0 & 99.2 & \textbf{6.9} & 93.7 & 8.6 & 95.2 \\
~ & ~ & 30 & 3.6 & 85.8 & 7.7 & 99.0 & \textbf{5.5} & 90.6 & 7.9 & 93.1 \\
\midrule
5 & 7 & 5 & 0.0 & 0.0 & \textbf{4.9} & 93.7 & 3.0 & 81.2 & 4.0 & 91.3 \\
~ & ~ & 15 & 0.0 & 0.0 & \textbf{5.0} & 93.9 & 4.2 & 90.9 & 4.7 & 89.2 \\
~ & ~ & 30 & 2.6 & 72.8 & \textbf{4.7} & 93.4 & 4.0 & 90.6 & 4.2 & 90.7 \\
\midrule
6 & 7 & 5 & 0.0 & 0.0 & 8.6 & 90.5 & \textbf{5.4} & 76.2 & 6.6 & 86.6 \\
~ & ~ & 15 & 0.0 & 0.0 & 9.0 & 91.1 & \textbf{6.8} & 83.7 & 8.0 & 85.0 \\
~ & ~ & 30 & \textbf{3.9} & 67.8 & 8.5 & 90.5 & 6.4 & 84.6 & 6.7 & 85.1 \\
\bottomrule
\end{tabular}
}\end{table}

\subsection{Language Models and Watermark Methods} 

\begin{figure}
    \centering
    \includegraphics[width=\textwidth]{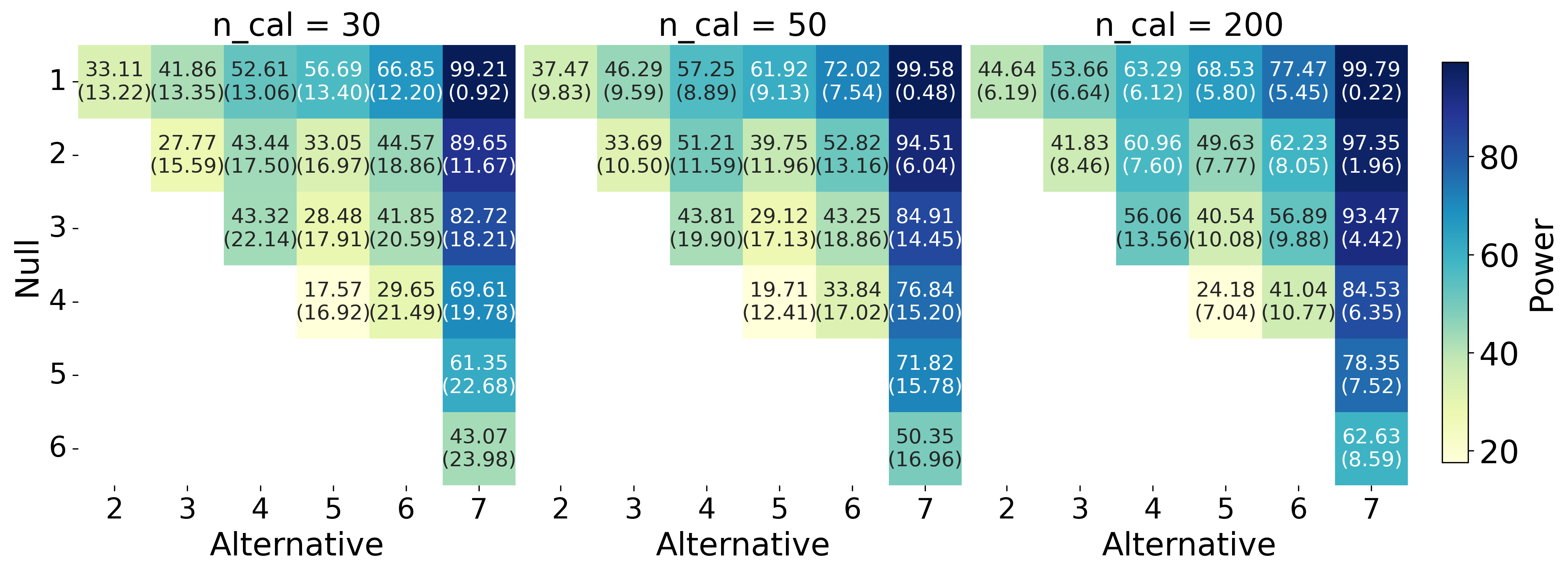}
    \includegraphics[width=\textwidth]{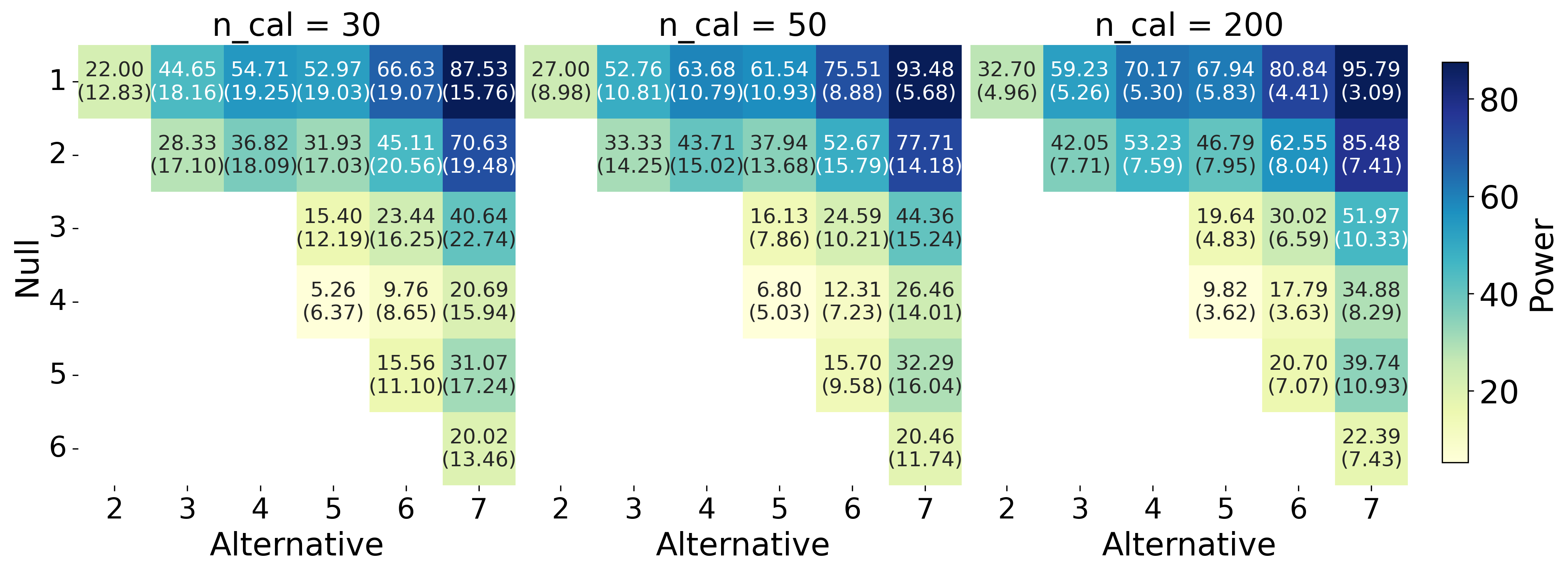}
    \caption{Detection Power. Top: Scenario 1, Standard Conformal Method using \texttt{Phi-4-mini-instruct} and Green/Red List watermark method; Bottom: Scenario 1, Standard Conformal Method using \texttt{Qwen2.5-7B-Instruct} and Green/Red List watermark method.}
    \label{fig:conformal_power_qwen}
\end{figure}

Different watermark methods will yield interactions that affect the detection power and FPR control. As shown in Figure~\ref{fig:conformal_power_qwen} (Top), for the \texttt{Phi-4-mini-instruct} model, when we use the Green/Red List watermark method, the detection power can be lower than using the Gumbel-Max watermark method. This is likely due to the lower detection power of the Green/Red List watermark method itself with the chosen parameters, as observed in the literature \citep{xie2024debiasing, fernandez2023three}. 

Moreover, different language models, even with the same watermark method, can yield different detection power. For example, as shown in Figure~\ref{fig:conformal_power_qwen} (Bottom), the \texttt{Qwen2.5-7B-Instruct} model has lower detection power than the \texttt{Phi-4-mini-instruct} model when using the Green/Red List watermark method. Since watermark signal strength depends on the text generation entropy \citep{kirchenbauer2023watermark,fernandez2023three}, when compared against the \texttt{Qwen2.5-7B-Instruct} model, \texttt{Phi-4-mini-instruct} model may have the tendency to generate more diverse text, leading to stronger watermark signals. This difference, however, presents a more challenging scenario for the instructor, as curating a calibration set with the same model as the test set may not be feasible in practice, given the vast number of choices of language models available in the market.

\section{Discussion}
\label{sec:discussion}

\subsection{Toward Post-AI Academic Integrity}

Traditional approaches to the detection of academic dishonesty have relied mainly on human judgment or technologically limited systems that do not capture the nuanced reality of contemporary classrooms with readily accessible AI models \citep{khalil2023will, casal2023can, fleckenstein2024teachers, gao2023comparing, rathi2024gpt, foltynek2019academic}. While recent advances in AI detection accuracy represent meaningful progress \citep{jiang2024detecting, campino2024unleashing, hyatt2025using, jiang2024towards, hyatt2025using}, there is a pronounced lack of granularity in design: these approaches primarily focus on distinguishing texts written entirely by humans or AI and overlook the more common spectrum of mixed authorship seen in modern classrooms \citep{jiang2024detecting, campino2024unleashing, hyatt2025using, jiang2024towards, hyatt2025using, fraser2025detecting}. Although some researchers investigate approaches to evade AI detection\citep{zhang2024detection, perkins2024genai, sadasivan2023can}, the findings fail to capture the nuances of mixed authorship, which dominates education in the post-AI era. Our integration of watermarking and conformal prediction techniques shed lights on a promising pathway toward embracing AI in education, wherein ethical human-AI collaboration is actively encouraged while simultaneously preventing over-reliance on AI tools \citep{eaton2025global, eaton2023postplagiarism, luo2024critical, perkins2024artificial, furze2024ai, mollick2024instructors}.

\subsection{Methodological Limitations and Future Research Directions}

Despite the encouraging results of our study, substantial opportunities remain for methodological and theoretical advancement. Our empirical analysis reveals that while robust detection power persists between grammar-checking AI edits and complete AI-generated content, significant gaps remain in detecting moderate AI guideline violations.

We identify two complementary research directions for methodological improvements. First, researchers can advance watermarking techniques to generate stronger statistical signals that correlate more precisely with AI involvement levels \citep{li2025statistical, zhao2024sok} and to maintain more consistent performance \citep{kirchenbauer2023watermark, aaronson, xie2024debiasing, he2025distributional}.
Our comparative analysis across different AI models and watermarking scheme reveals noticeable variability in detection efficacy. Moreover, though not discussed in our paper, existing watermarking remains vulnerable to adversarial attacks and demands improvement\citep{kirchenbauerreliability, krishna2023paraphrasing, kuditipudi2023robust}.

The second research front involves improving conformal prediction methodologies. The conservative nature of our proposed methods, while prioritizing FPR control, may compromise statistical power in certain scenarios \citep{angelopoulos2024theoretical}. Recent developments in conformal prediction theory offer promising avenues for achieving superior FPR-power trade-offs \citep{vovk2015cross, barber2023conformal, marandon2024adaptive}. Furthermore, the three educational scenarios examined in this study, while representative of numerous classroom configurations, are not comprehensive. For example, curating a calibration dataset that can accommodate a diverse range of available AI tools and student demographics remains a significant challenge \citep{stone2023student, Coldwell2024AICheating, stone2024generative}. Moreover, as student behavior changes in the post-ChatGPT era, historical calibration data may become insufficiently representative of contemporary AI usage patterns, introducing distribution shifts that demand adaptive online conformal methodologies \citep{bhatnagar2023improved, gibbs2024conformal, zhou2024conformalized}. Future research should explore adaptive techniques that can dynamically adjust to evolving classroom settings while maintaining statistical rigor.

Our simulation framework, while demonstrating the technological promise of classroom watermarking applications, may not fully capture the complexity of student-AI interactions in reality. Our experimental design applies AI editing to completed essays in a single iteration, whereas students engage iteratively with AI-powered tools such as Grammarly \citep{fitria2021grammarly} throughout the writing process. The cumulative impact of these complex interactive patterns on watermark signal and subsequent detection accuracy will require future empirical investigation. 

\subsection{Statistical Rigor and Practical Implementation}

In this work, we emphasize the importance of FPR control for AI detection systems. The statistical guarantees that constrain the probability of false accusations can limit the subjectivity that have historically characterized academic integrity enforcement \citep{stone2023student, Coldwell2024AICheating, stone2024generative}. Moreover, the psychological and academic consequences of false accusation, particularly for vulnerable student populations, demand frameworks that prioritize fairness and accuracy.

However, we emphasize that statistical significance does not automatically imply practical significance. Educators should employ our framework as one component of comprehensive assessment strategies that incorporate multiple detection methods and consider individual student academic histories and circumstances \citep{moorhouse2023generative, mcdonald2025generative}.

\subsection{Organizational Implementation Challenges}

Realistic deployment of watermark-based detection systems requires substantial organizational coordination and institutional commitment. While watermarking techniques have been implemented in commercial AI models such as Google's Gemini \citep{dathathri2024scalable}, widespread adoption remains limited due to policy uncertainties and technical implementation challenges \citep{rijsbosch2025adoption, zhao2024sok}.

On the other hand, all conformal prediction methods require historical calibration data, and while our scenarios are presented from individual educator perspectives, the collection and maintenance of representative calibration datasets likely necessitate collaboration between institutional information technology departments and pedagogical experts. Moreover, the computational overhead associated with collecting the calibration datasets in our framework may exceed the capacity of a single educator or institution, suggesting that scalable commercial solutions may prove more practical.

The successful deployment of such systems ultimately depends on the development of clear institutional policies regarding AI-use in classrooms, AI-literacy training for educators, and robust technical infrastructure capable of supporting large-scale detection. These implementation challenges, while substantial, do not diminish the theoretical and practical promise of our proposed methodology for maintaining academic integrity in the post-AI classrooms.
\subsection*{Data Repository/Code}
The code for reproducing experiments is available online at \url{https://github.com/Xieyangxinyu/Watermark-in-the-Classroom}.
\subsection*{Acknowledgments}

The authors would like to thank Xiang Li, Qi Long, Tanwi Mallick and Linda H. Zhao for helpful comments in earlier drafts of this paper. The authors extend their appreciation to the Mack Institute for Innovation Management at the Wharton School, University of Pennsylvania, for fostering an interdisciplinary research environment that facilitated the development of this methodological framework.

\subsection*{Disclosure Statement}

The authors acknowledge financial support from the Mack Institute Research Fellowship. This research used computational resources provided by the Argonne Leadership Computing Facility, under Contract No. DE-AC02-06CH11357. W.~J.~S.~is supported by NSF DMS-2310679, a Meta Faculty Research Award and Wharton AI for Business. Z.~R.~is supported by the National Science Foundation (NSF) under grant DMS-2413135
and Wharton Analytics.

\bibliographystyle{plainnat}
\bibliography{references}

\appendix
\appendix

\section{Illustration Figures: Further Details}

\cref{fig:pval_dist_grammar,fig:visual_inspection,fig:conformal_pvals,fig:dist_shift,fig:bleu_dist} are generated by simulating the respective detection methods on a subset of the ETS corpus. Specifically, this subset includes essays responding to the following writing prompt:

``Do you agree or disagree with the following statement? It is better to have broad knowledge of many academic subjects than to specialize in one specific subject.''

All AI-edits shown in these figures are made by the \texttt{Phi-4-mini-instruct} model using the Gumbel-Max watermarking method. Figure \ref{fig:pval_dist_grammar} is generated using 200 randomly sampled essays from this subset, comparing the original human-written essays against the same essays edited by using the first instruction prompt defined in \ref{sec:ai_assistance}. \cref{fig:visual_inspection,fig:bleu_dist} are generated using 100 essays, 50 of which are edited using the first instruction prompt defined in \ref{sec:ai_assistance}, while the remaining 50 are edited by a uniformly randomly chosen instruction prompt (other than the first). Figure~\ref{fig:conformal_pvals} is generated using 50 calibration essays and 1024 test essays, while Figure \ref{fig:hierarchical_pvals} is generated using 200 calibration essays and 185 test essays. All test essays are edited in two ways:
\begin{itemize}
    \item Inliers: Each essay is edited using the first instruction prompt defined in Section~\ref{sec:ai_assistance}.
    \item Outliers: Each essay is edited by a uniformly randomly chosen instruction prompt (other than the first), then filtered using the definition introduced in Section \ref{sec:refine_def}. Figure~\ref{fig:conformal_pvals} contains 565 outliers while Figure \ref{fig:hierarchical_pvals} contains 103 outliers.
\end{itemize}

Figure \ref{fig:dist_shift} is generated by comparing all essays in the LOCNESS corpus and 200 essays in the ETS corpus subset, all of which are edited using the first instruction prompt defined in Section~\ref{sec:ai_assistance}. 

\section{Methods: Further Details}

\subsection{Details of Datasets, Language Models and Watermarking Schemes}
\label{sec:dataset_LM_watermark}
Our experiments use two publicly available corpora:
\begin{itemize}
  \item \textbf{ETS TOEFL11 (ETS) Corpus} \citep{blanchard2013toefl11}: We include 8,884 essays by non-native speakers across a range of language backgrounds and eight different writing prompts, each within 250–400 words (mean 321.3, SD 36.3). 
  \item \textbf{LOCNESS Corpus} \citep{granger1998computer}: We sample 385 essays by native speakers (American and British university students + A-level examinees), truncated to 250–400 words (mean 354.9, SD 41.1). 273 unique essay identifiers (topics) are represented in this sample. This dataset contains 171 essays spanning 59 different topics with multiple essays per topic, while the remaining 214 essays each represent a unique topic.
\end{itemize}
Language models are configured with temperature parameter $\tau = 0.7$. Both watermarking schemes use a context window size $n = 4$. For Green/Red List watermark, we set hyperparameters $\delta = 2$ and $\gamma = 0.5$.
All experiments were conducted locally using NVIDIA A100-SXM4-40GB GPUs, each with 40GB of VRAM.

\subsection{Implementation Details of Quantile Shift}
\label{sec:quantile_shift}
The method first calculates a specified lower quantile (by default, the $\alpha$-quantile) of the scores from both the calibration set (the LOCNESS corpus and the $m$ sampled non-native essays) and the minority set (only the $m$ non-native essays). The shift is computed as the difference between these quantiles. When the sample size of the minority set is small (i.e., $\leq \frac{1}{2\alpha}$), the quantile is replaced with the minimum score to ensure robustness; when the size is moderately small (i.e., $\leq \frac{1}{\alpha}$), a more conservative quantile of level $2\alpha$ is used. The density $\hat q$ is then estimated by shifting $\hat p$ using this quantile difference and rescaling with the standard deviations of both sets. This approach allows for a better approximation of $\hat q$ in cases where the score distributions differ more significantly in the lower tail than in the mean.

\begin{algorithm}
\caption{Quantile Shift Density Estimation}
\begin{algorithmic}[1]
\Require Scores from majority class $p\_scores$, minority class $q\_scores$, significance level $\alpha$
\Ensure Estimated density ratio $\hat{q}(x)$ via transformation of $\hat{p}(x)$

\State Compute KDE estimate $\hat{p}(x)$ from $p\_scores$
\State Compute $\sigma_p \gets \mathrm{std}(p\_scores)$, $\sigma_q \gets \mathrm{std}(q\_scores)$

\If{$|q\_scores| \leq \frac{1}{2\alpha}$}
    \State $q_\alpha \gets \min(q\_scores)$
    \State $p_\alpha \gets \mathrm{quantile}(p\_scores, \frac{1}{|q\_scores|})$
\ElsIf{$|q\_scores| \leq \frac{1}{\alpha}$}
    \State $q_\alpha \gets \mathrm{quantile}(q\_scores, 2\alpha)$
    \State $p_\alpha \gets \mathrm{quantile}(p\_scores, 2\alpha)$
\Else
    \State $q_\alpha \gets \mathrm{quantile}(q\_scores, \alpha)$
    \State $p_\alpha \gets \mathrm{quantile}(p\_scores, \alpha)$
\EndIf

\State Define $\hat{q}(x) \gets \hat{p}\left( \frac{(x - q_\alpha)}{\sigma_q} \cdot \sigma_p + p_\alpha \right)$
\State \Return $\hat{q}(x)$
\end{algorithmic}
\end{algorithm}

\subsection{Refining the Definition of Guideline Violation}
\label{sec:refine_def}
There is an important nuance in AI-usage guideline violations: not all violations carry the same degree of impact on academic integrity. Consider a common scenario: a student independently produces a well-written essay, then—against course guidelines—uses AI to polish the final draft with different prompts. In such cases, while technically a violation has occurred, the resulting essay contains relatively little AI-generated content. To quantify this phenomenon, we measure the similarity between original essays and their AI-edited versions using BLEU scores \citep{papineni2002bleu}. Higher BLEU scores indicate minimal modification by AI, while lower scores suggest substantial AI contribution. Figure~\ref{fig:bleu_dist_a} illustrates how essays with different degrees of AI modification distribute across the watermarking detection space. This observation led us to refine our definition of AI-usage guideline violations into two categories: 

\begin{figure}
    \centering
    \begin{subfigure}[b]{0.48\textwidth}
        \centering
        \includegraphics[width=\textwidth]{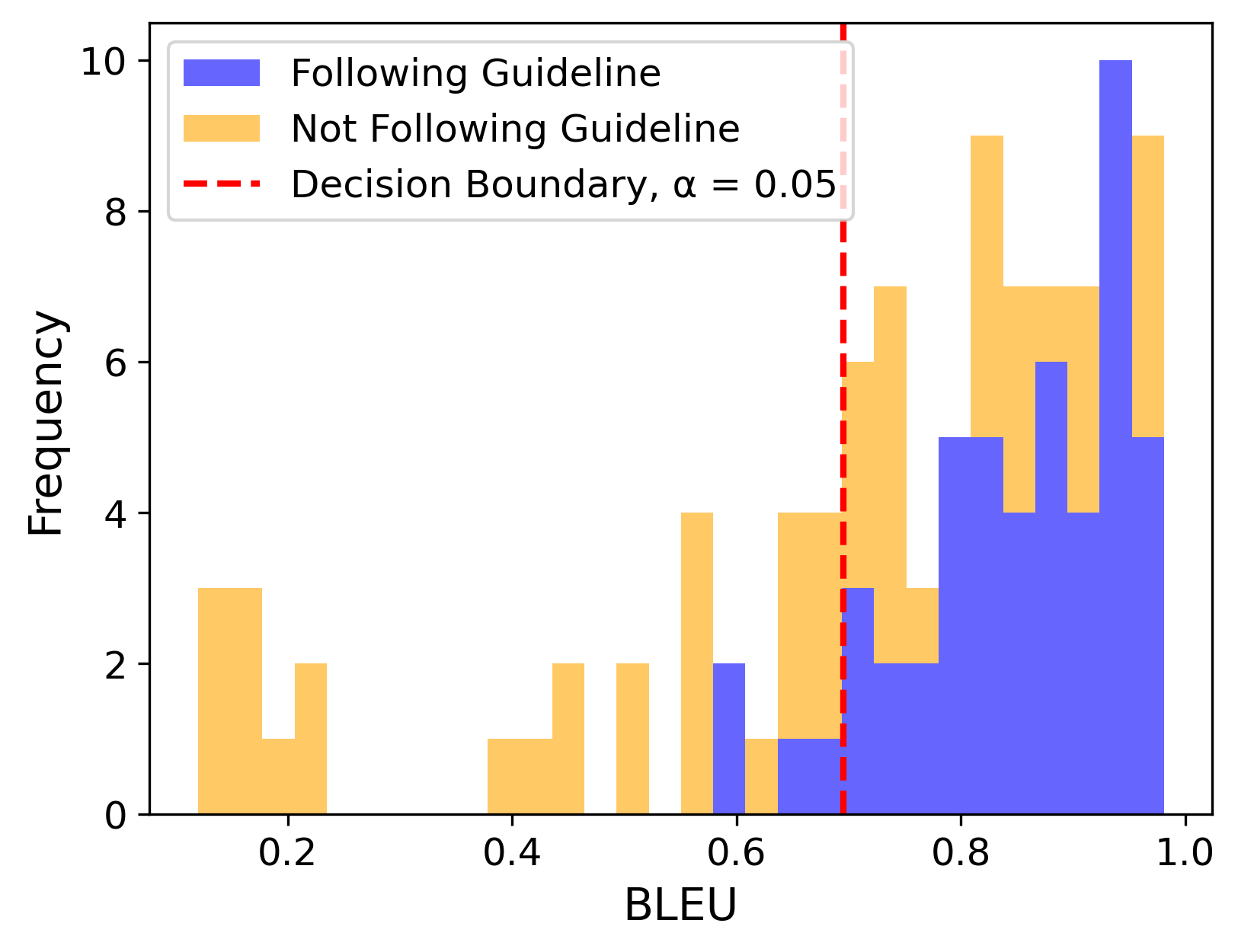}
        \caption{Distribution by BLEU score}
        \label{fig:bleu_dist_a}
    \end{subfigure}
    \hfill
    \begin{subfigure}[b]{0.48\textwidth}
        \centering
        \includegraphics[width=\textwidth]{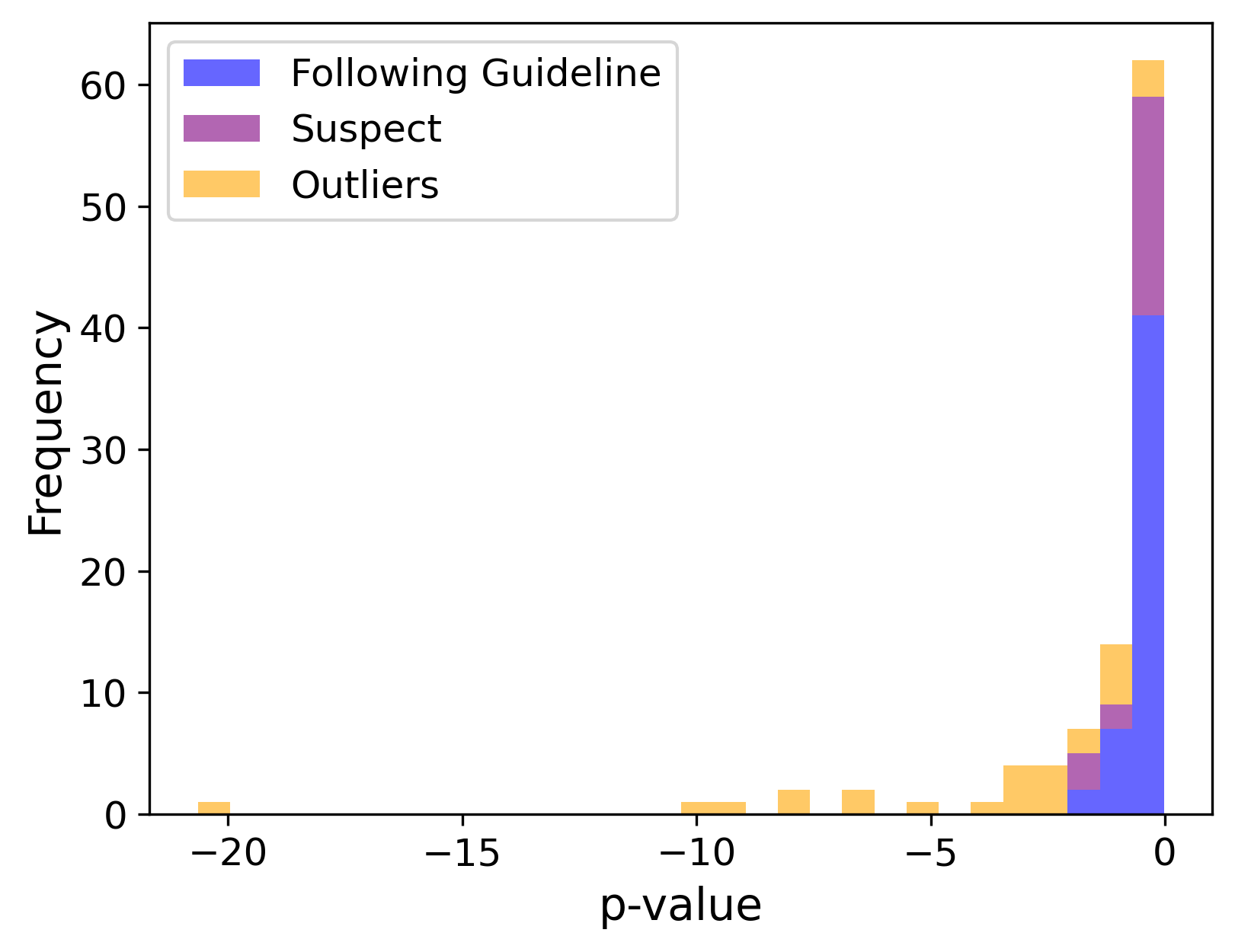}
        \caption{Distribution by watermarking p-value}
        \label{fig:bleu_dist_b}
    \end{subfigure}
    \caption{Complementary views of AI assistance categorization. 100 sample essays from the ETS corpus are included, 50 of which are AI-edited using the first prompt (i.e., prompt 1 in Section~\ref{sec:ai_assistance}) and the other 50 using the remaining six prompts (i.e., prompts 2–7 uniformly randomly). (a) BLEU score distribution comparing original essays to their AI-edited versions, where higher scores indicate minimal AI modification. (b) Watermarking p-value distribution (on $\log_{100}$ scale) showing the same data categorized into three groups: essays following guidelines (blue), clear violations with substantial AI content (yellow), and ``suspect'' cases (purple) where students used AI beyond permitted grammar checking but with limited content modification. }
    \label{fig:bleu_dist}
\end{figure}

\begin{itemize}
    \item \textbf{Outliers}: a student violates AI-usage guidelines and the result leads to significant changes in the essay, as measured by a low BLEU score. Specifically, we classify an edited essay as a clear outlier if it satisfy both conditions: (1) the resulting BLEU score by following the AI-usage guidelines is higher than that by violating the guidelines, and (2) the resulting BLEU score by violating the guidelines is below the $\alpha = 0.05$ quantile of the test dataset (explained below).
    
    \item \textbf{Suspect}: a student violates AI-usage guidelines but the result leads to only minor changes in the essay (high BLEU score). In other words, alternative AI-edits are applied, but the resulting essay is not classified as an outlier.
\end{itemize}

\subsubsection{Implementation Detail of the BLEU Score} The method tokenizes both the reference and candidate texts into lowercased word tokens. BLEU is then computed by comparing the candidate against the reference using modified n-gram precisions. In our implementation, it uses a weighted average of unigram and bigram precisions, with default weights of (0.5, 0.5), placing equal emphasis on both. 

\section{Additional Results}
\begin{figure}
\centering
\includegraphics[width=0.95\textwidth]{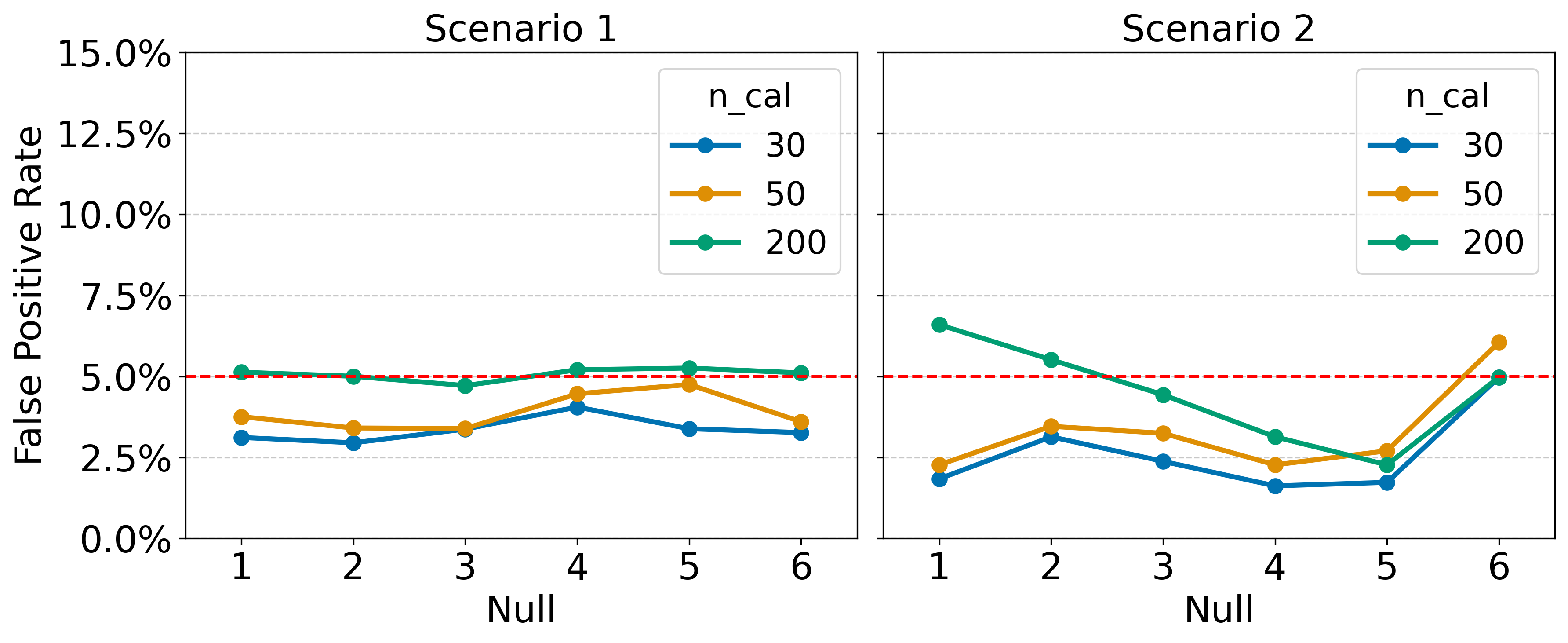}
\caption{Average FPR in Scenario 1 and Scenario 2. Model: \texttt{Phi-4-mini-instruct}, Watermarking Method: \texttt{Green/Red List}.}
\label{fig:fpr_phi_maryland}
\end{figure}

\begin{figure}
\centering
\includegraphics[width=0.95\textwidth]{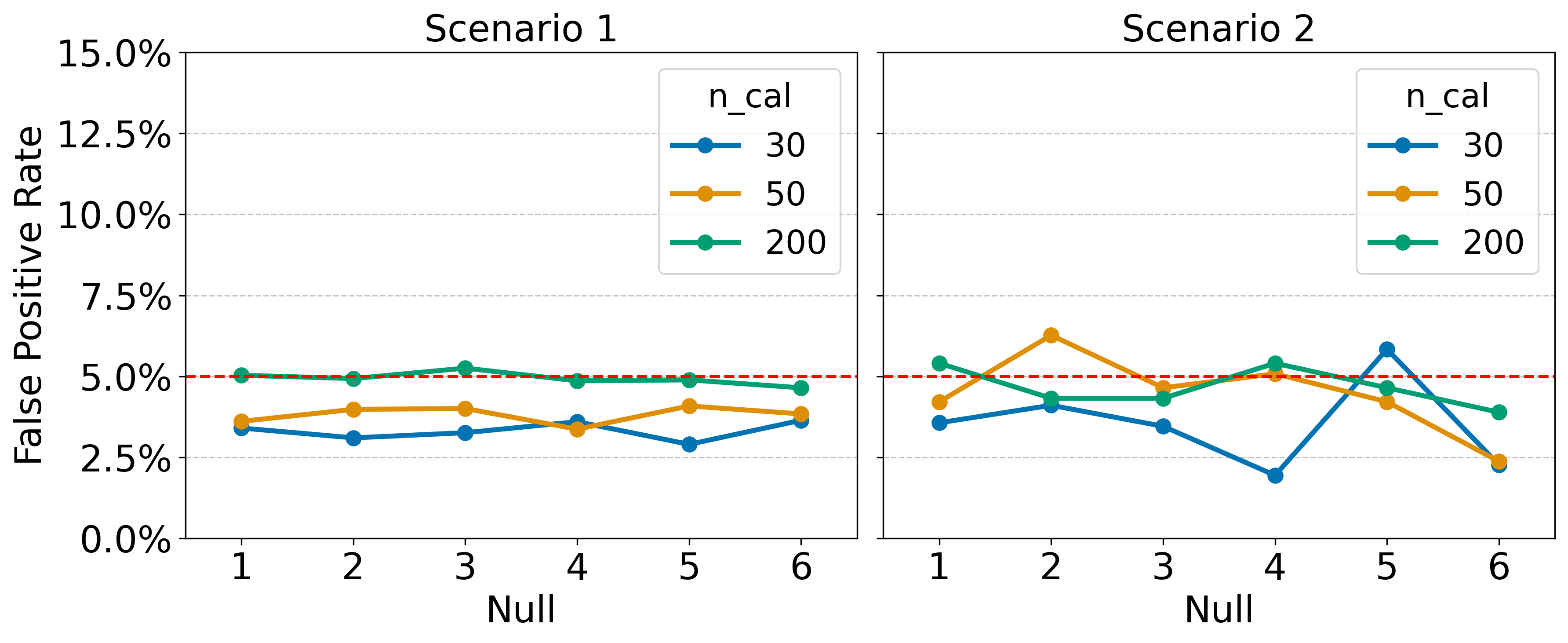}
\caption{Average FPR in Scenario 1 and Scenario 2. Model: \texttt{Qwen2.5-7B-Instruct}, Watermarking Method: \texttt{Gumbel-Max}.}
\label{fig:FPR_qwen_Gumbel}
\end{figure}

\begin{figure}
\centering
\includegraphics[width=0.95\textwidth]{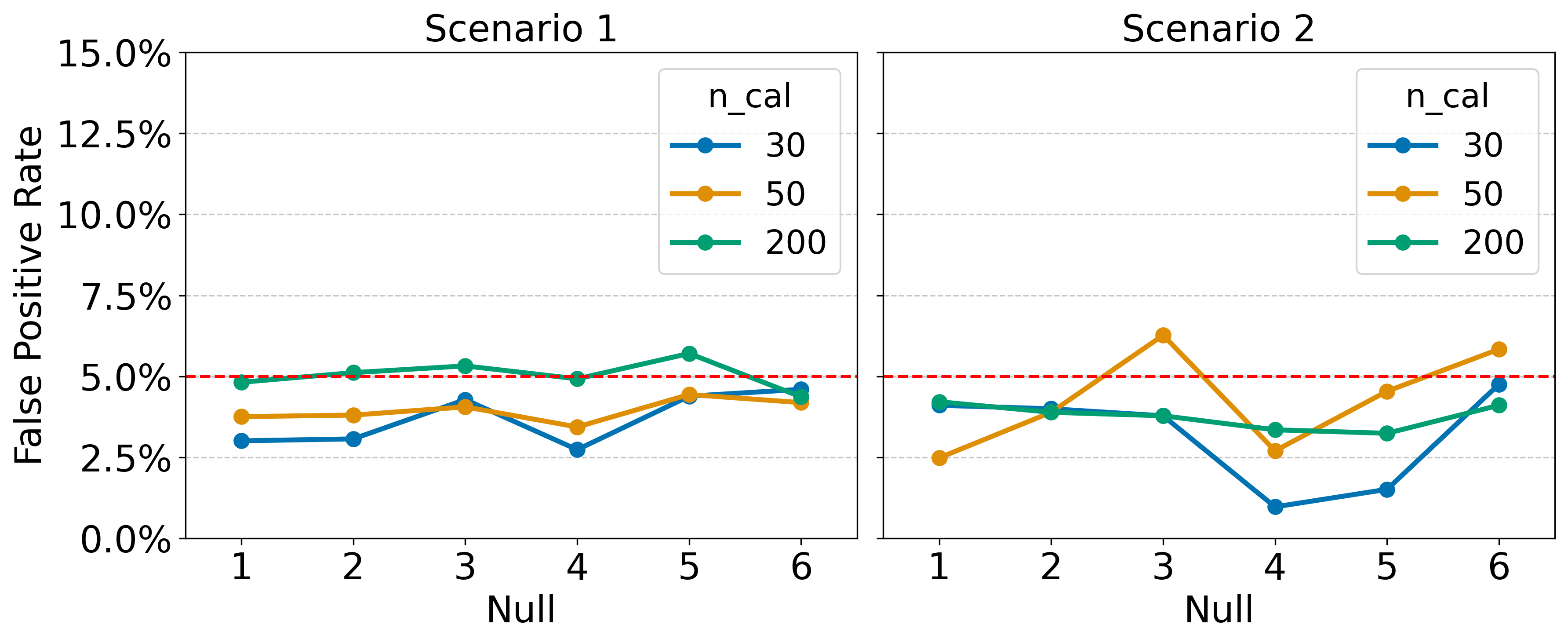}
\caption{Average FPR in Scenario 1 and Scenario 2. Model: \texttt{Qwen2.5-7B-Instruct}, Watermarking Method: \texttt{Green/Red List}.}
\label{fig:fpr_qwen_maryland}
\end{figure}

\begin{figure}
    \centering
    \includegraphics[width=\textwidth]{figs/ETS_conformal_phi_maryland_power_base_alternative.png}
    \includegraphics[width=\textwidth]{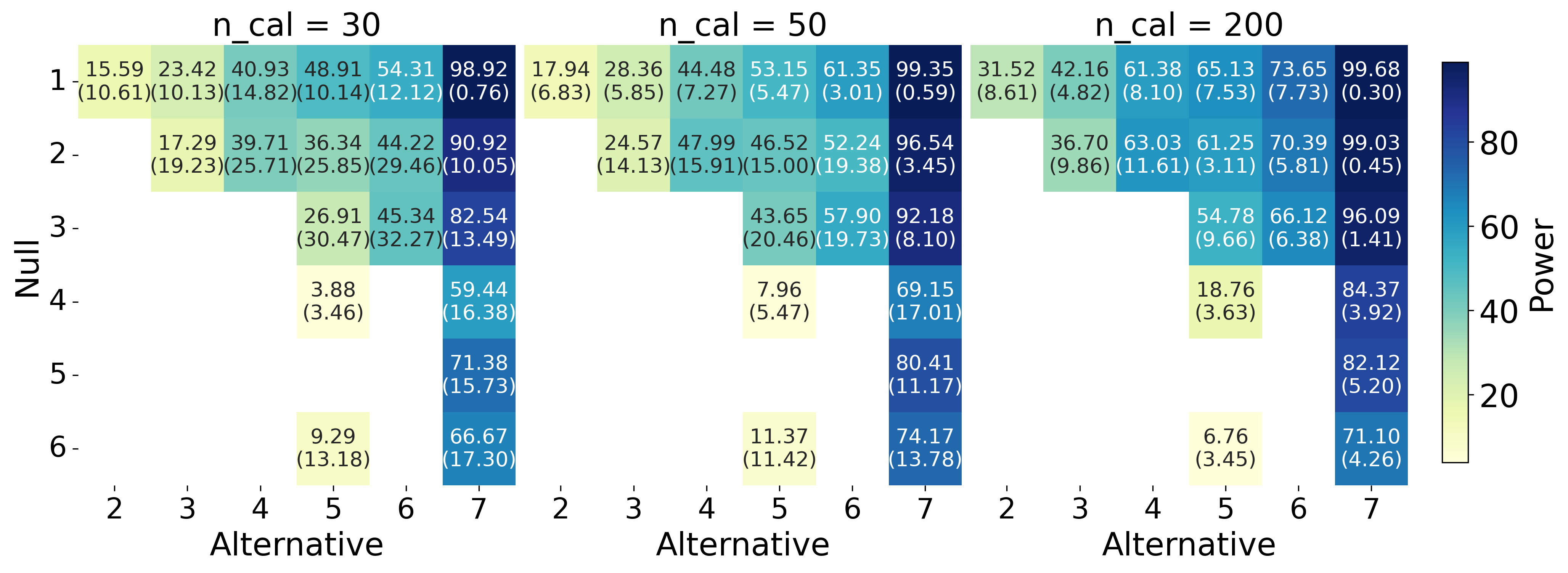}
    \caption{Detection Power. Top: Scenario 1, Standard Conformal Method. (ETS); Bottom: Scenario 2 (LOCNESS), Hierarchical Conformal Method. Model: \texttt{Phi-4-mini-instruct}, Watermarking Method: \texttt{Green/Red List}.}
    \label{fig:conformal_phi_maryland}
\end{figure}

\begin{figure}
    \centering
    \includegraphics[width=\textwidth]{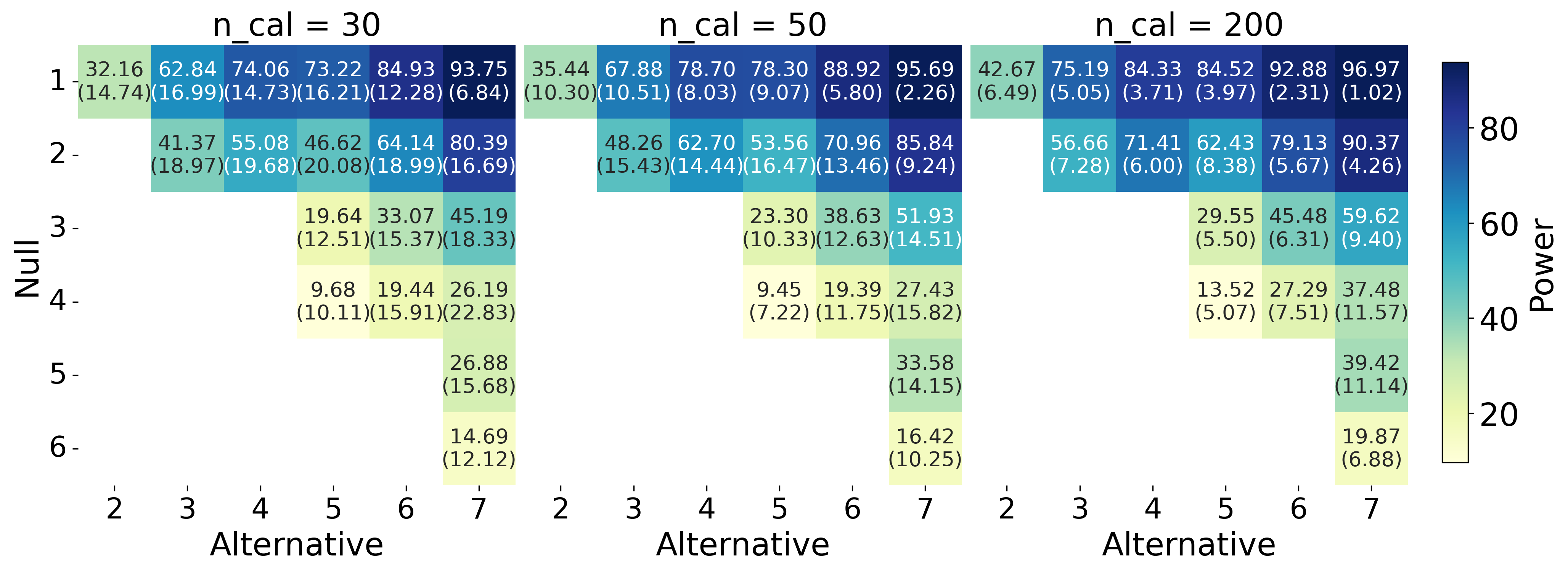}
    \includegraphics[width=\textwidth]{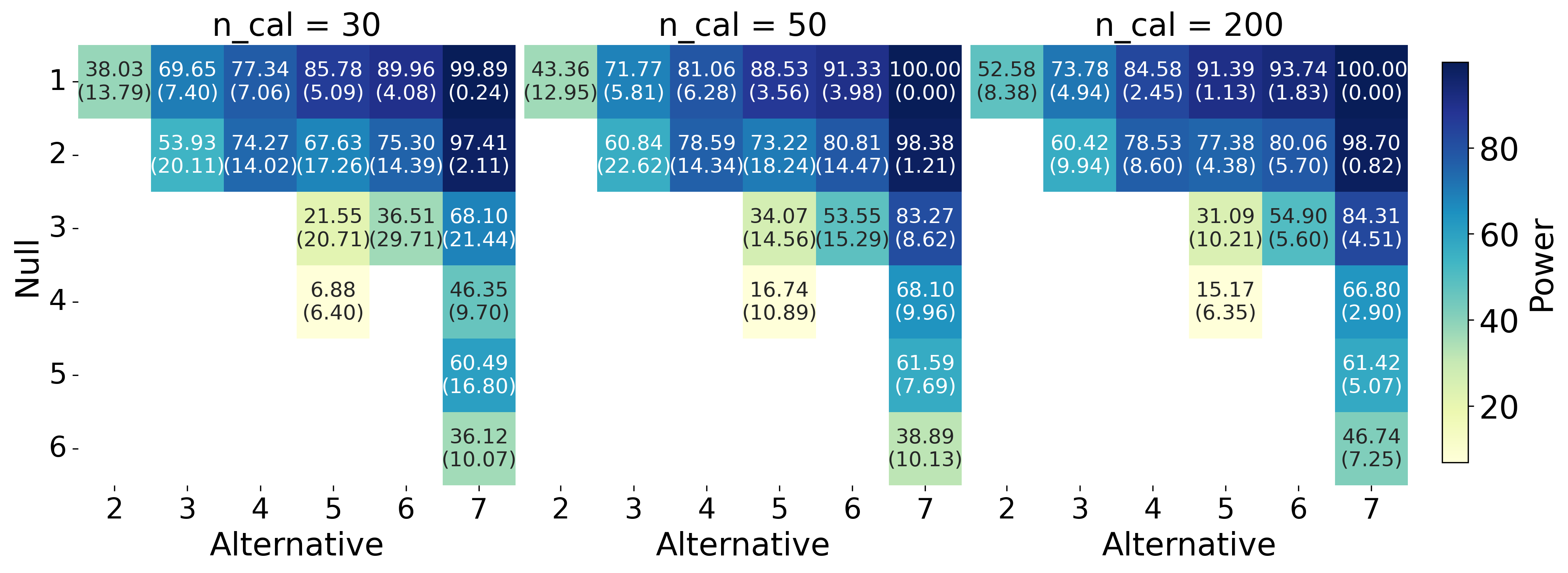}
    \caption{Detection Power. Top: Scenario 1 (ETS), Standard Conformal Method. Bottom: Scenario 2 (LOCNESS), Hierarchical Conformal Method. Model: \texttt{Qwen2.5-7B-Instruct}, Watermarking Method: \texttt{Gumbel-Max}.}
    \label{fig:conformal_qwen_openai}
\end{figure}

\begin{figure}
    \centering
    \includegraphics[width=\textwidth]{figs/ETS_conformal_qwen_maryland_power_base_alternative.png}
    \includegraphics[width=\textwidth]{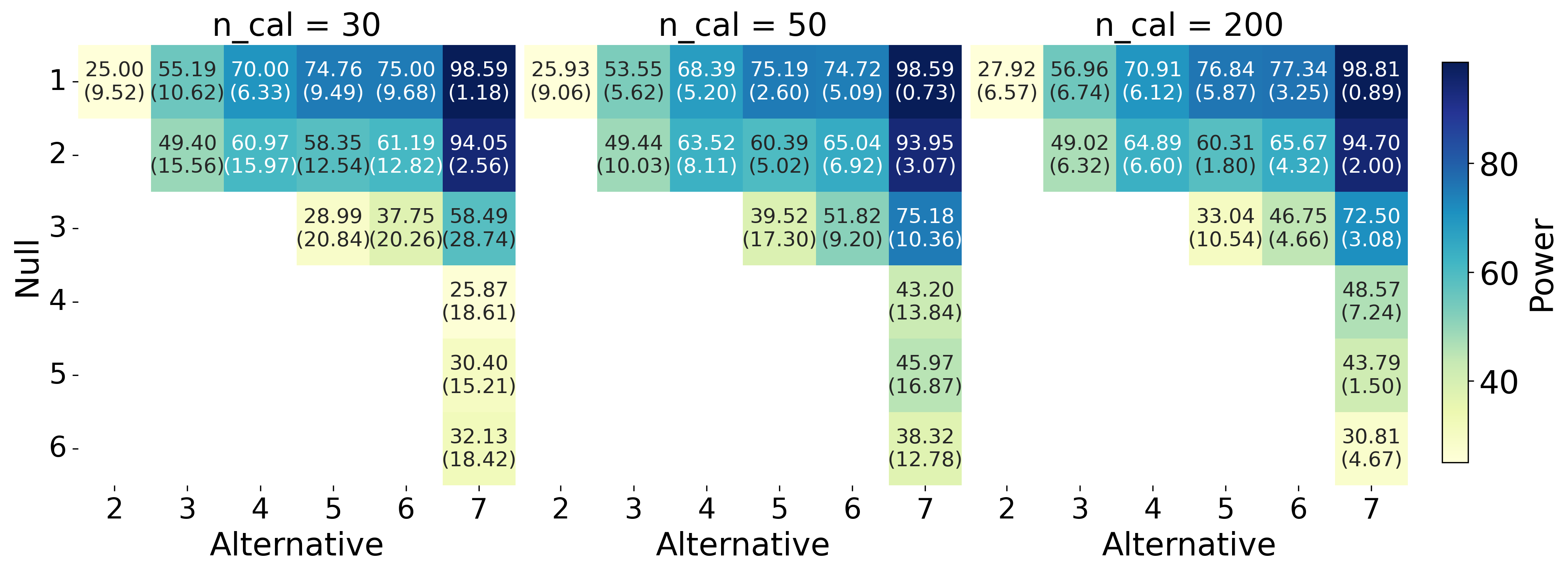}
    \caption{Detection Power. Top: Scenario 1 (ETS), Standard Conformal Method. Bottom: Scenario 2 (LOCNESS), Hierarchical Conformal Method. Model: \texttt{Qwen2.5-7B-Instruct}, Watermarking Method: \texttt{Green/Red List}.}
    \label{fig:conformal_qwen_maryland}
\end{figure}

Figure~\ref{fig:fpr_phi_maryland}, Figure~\ref{fig:FPR_qwen_Gumbel}, and Figure~\ref{fig:fpr_qwen_maryland} plot the FPR control of the standard conformal method and the hierarchical conformal method across different language models and watermarking methods. Similarly, Figure~\ref{fig:conformal_phi_maryland}, Figure~\ref{fig:conformal_qwen_openai}, and
Figure~\ref{fig:conformal_qwen_maryland} present comparisons of the detection power achieved under these methods and configurations.

\begin{table}[ht]
\centering
\caption{Mapping of tables to language models and watermarking methods}
\label{tab:appendix_table_mapping}
\begin{tabular}{lll}
\hline
\textbf{Table Reference} & \textbf{Language Model} & \textbf{Watermarking Method} \\
\hline
\ref{tab:phi_openai_n5_ETS_conformal_weighted}, \ref{tab:phi_openai_n15_ETS_conformal_weighted}, \ref{tab:phi_openai_n30_ETS_conformal_weighted} & \texttt{Phi-4-mini-instruct} & \texttt{Gumbel-Max} \\
\ref{tab:phi_maryland_n5_ETS_conformal_weighted}, \ref{tab:phi_maryland_n15_ETS_conformal_weighted}, \ref{tab:phi_maryland_n30_ETS_conformal_weighted} & \texttt{Phi-4-mini-instruct} & \texttt{Green/Red List} \\
\ref{tab:qwen_openai_n5_ETS_conformal_weighted}, \ref{tab:qwen_openai_n15_ETS_conformal_weighted}, \ref{tab:qwen_openai_n30_ETS_conformal_weighted} & \texttt{Qwen2.5-7B-Instruct} & \texttt{Gumbel-Max} \\
\ref{tab:qwen_maryland_n5_ETS_conformal_weighted}, \ref{tab:qwen_maryland_n15_ETS_conformal_weighted}, \ref{tab:qwen_maryland_n30_ETS_conformal_weighted} & \texttt{Qwen2.5-7B-Instruct} & \texttt{Green/Red List} \\
\hline
\end{tabular}
\end{table}
We report the complete results on the performance of the three conformal prediction methods under varying NNES calibration sizes $m\in\{5, 15, 30\}$. Table~\ref{tab:appendix_table_mapping} provides an overview of the mapping between the tables, language models, and watermarking methods.

\begin{table}[h!]
\tiny
\centering
\caption{Weighted Conformal Methods: FPR and Power for Different Calibration Sizes and AI-Usage Guidelines. Model: \texttt{Phi-4-mini-instruct}, Method: \texttt{Gumbel-Max}, NNES calibration
size $m$: 5.}
\label{tab:phi_openai_n5_ETS_conformal_weighted}
\resizebox{\textwidth}{!}{%
\begin{tabular}{c c | c c | c c | c c | c c }
\toprule
Null & Alt. & \multicolumn{2}{c|}{In Dist. Only} & \multicolumn{2}{c|}{Comb. Unweighted} & \multicolumn{4}{c}{Comb. Weighted}\\
& & & & & & \multicolumn{2}{c|}{Quantile} & \multicolumn{2}{c}{Mean} \\
& & FPR & Power & FPR & Power & FPR & Power & FPR & Power \\
\midrule
1 & 2 & 0.0 & 0.0 & 12.7 & 69.5 & 6.4 & 53.6 & 9.7 & 63.4 \\
~ & 3 & ~ & 0.0 & ~ & 77.2 & ~ & 63.2 & ~ & 71.8 \\
~ & 4 & ~ & 0.0 & ~ & 83.1 & ~ & 71.4 & ~ & 78.7 \\
~ & 5 & ~ & 0.0 & ~ & 89.6 & ~ & 79.7 & ~ & 86.0 \\
~ & 6 & ~ & 0.0 & ~ & 93.9 & ~ & 86.7 & ~ & 91.6 \\
~ & 7 & ~ & 0.0 & ~ & 100.0 & ~ & 100.0 & ~ & 100.0 \\
\midrule
2 & 3 & 0.0 & 0.0 & 13.2 & 80.9 & 5.8 & 55.7 & 10.1 & 74.1 \\
~ & 4 & ~ & 0.0 & ~ & 91.2 & ~ & 71.4 & ~ & 86.8 \\
~ & 5 & ~ & 0.0 & ~ & 87.3 & ~ & 62.4 & ~ & 81.1 \\
~ & 6 & ~ & 0.0 & ~ & 93.3 & ~ & 74.4 & ~ & 89.2 \\
~ & 7 & ~ & 0.0 & ~ & 100.0 & ~ & 99.4 & ~ & 99.9 \\
\midrule
3 & 5 & 0.0 & 0.0 & 9.4 & 79.9 & 4.7 & 49.0 & 7.5 & 70.6 \\
~ & 6 & ~ & 0.0 & ~ & 89.9 & ~ & 65.3 & ~ & 83.8 \\
~ & 7 & ~ & 0.0 & ~ & 99.8 & ~ & 97.3 & ~ & 99.6 \\
\midrule
4 & 5 & 0.0 & 0.0 & 8.0 & 54.3 & 4.4 & 29.2 & 7.1 & 48.3 \\
~ & 6 & ~ & 0.0 & ~ & 74.0 & ~ & 45.8 & ~ & 68.3 \\
~ & 7 & ~ & 0.0 & ~ & 99.1 & ~ & 80.5 & ~ & 97.3 \\
\midrule
5 & 7 & 0.0 & 0.0 & 4.9 & 93.7 & 3.0 & 81.2 & 4.0 & 91.3 \\
\midrule
6 & 7 & 0.0 & 0.0 & 8.6 & 90.5 & 5.4 & 76.2 & 6.6 & 86.6 \\
\bottomrule
\end{tabular}
}\end{table}
\begin{table}[h!]
\tiny
\centering
\caption{Weighted Conformal Methods: FPR and Power for Different Calibration Sizes and AI-Usage Guidelines. Model: \texttt{Phi-4-mini-instruct}, Method: \texttt{Gumbel-Max}, NNES calibration
size $m$: 15.}
\label{tab:phi_openai_n15_ETS_conformal_weighted}
\resizebox{\textwidth}{!}{%
\begin{tabular}{c c | c c | c c | c c | c c }
\toprule
Null & Alt. & \multicolumn{2}{c|}{In Dist. Only} & \multicolumn{2}{c|}{Comb. Unweighted} & \multicolumn{4}{c}{Comb. Weighted}\\
& & & & & & \multicolumn{2}{c|}{Quantile} & \multicolumn{2}{c}{Mean} \\
& & FPR & Power & FPR & Power & FPR & Power & FPR & Power \\
\midrule
1 & 2 & 0.0 & 0.0 & 12.7 & 69.4 & 8.4 & 60.0 & 7.9 & 58.7 \\
~ & 3 & ~ & 0.0 & ~ & 77.1 & ~ & 69.4 & ~ & 68.1 \\
~ & 4 & ~ & 0.0 & ~ & 82.9 & ~ & 76.7 & ~ & 75.9 \\
~ & 5 & ~ & 0.0 & ~ & 89.5 & ~ & 84.1 & ~ & 83.2 \\
~ & 6 & ~ & 0.0 & ~ & 93.8 & ~ & 90.3 & ~ & 89.6 \\
~ & 7 & ~ & 0.0 & ~ & 100.0 & ~ & 100.0 & ~ & 100.0 \\
\midrule
2 & 3 & 0.0 & 0.0 & 13.0 & 80.7 & 6.9 & 59.9 & 7.9 & 63.0 \\
~ & 4 & ~ & 0.0 & ~ & 91.1 & ~ & 74.2 & ~ & 77.2 \\
~ & 5 & ~ & 0.0 & ~ & 87.1 & ~ & 66.6 & ~ & 69.6 \\
~ & 6 & ~ & 0.0 & ~ & 93.1 & ~ & 77.5 & ~ & 80.1 \\
~ & 7 & ~ & 0.0 & ~ & 100.0 & ~ & 99.5 & ~ & 99.6 \\
\midrule
3 & 5 & 0.0 & 0.0 & 9.6 & 80.5 & 6.8 & 64.2 & 6.5 & 61.6 \\
~ & 6 & ~ & 0.0 & ~ & 90.3 & ~ & 77.6 & ~ & 75.2 \\
~ & 7 & ~ & 0.0 & ~ & 99.8 & ~ & 98.5 & ~ & 97.8 \\
\midrule
4 & 5 & 0.0 & 0.0 & 8.0 & 54.3 & 6.9 & 46.4 & 8.6 & 55.2 \\
~ & 6 & ~ & 0.0 & ~ & 73.9 & ~ & 64.6 & ~ & 72.5 \\
~ & 7 & ~ & 0.0 & ~ & 99.2 & ~ & 93.7 & ~ & 95.2 \\
\midrule
5 & 7 & 0.0 & 0.0 & 5.0 & 93.9 & 4.2 & 90.9 & 4.7 & 89.2 \\
\midrule
6 & 7 & 0.0 & 0.0 & 9.0 & 91.1 & 6.8 & 83.7 & 8.0 & 85.0 \\
\bottomrule
\end{tabular}
}\end{table}
\begin{table}[h!]
\tiny
\centering
\caption{Weighted Conformal Methods: FPR and Power for Different Calibration Sizes and AI-Usage Guidelines. Model: \texttt{Phi-4-mini-instruct}, Method: \texttt{Gumbel-Max}, NNES calibration
size $m$: 30.}
\label{tab:phi_openai_n30_ETS_conformal_weighted}
\resizebox{\textwidth}{!}{%
\begin{tabular}{c c | c c | c c | c c | c c }
\toprule
Null & Alt. & \multicolumn{2}{c|}{In Dist. Only} & \multicolumn{2}{c|}{Comb. Unweighted} & \multicolumn{4}{c}{Comb. Weighted}\\
& & & & & & \multicolumn{2}{c|}{Quantile} & \multicolumn{2}{c}{Mean} \\
& & FPR & Power & FPR & Power & FPR & Power & FPR & Power \\
\midrule
1 & 2 & 3.4 & 41.0 & 11.8 & 68.1 & 6.6 & 54.2 & 6.7 & 55.3 \\
~ & 3 & ~ & 52.5 & ~ & 76.1 & ~ & 64.4 & ~ & 65.4 \\
~ & 4 & ~ & 61.7 & ~ & 82.0 & ~ & 72.1 & ~ & 73.0 \\
~ & 5 & ~ & 70.1 & ~ & 88.8 & ~ & 79.9 & ~ & 80.9 \\
~ & 6 & ~ & 79.2 & ~ & 93.5 & ~ & 87.3 & ~ & 88.1 \\
~ & 7 & ~ & 99.9 & ~ & 100.0 & ~ & 100.0 & ~ & 100.0 \\
\midrule
2 & 3 & 3.5 & 44.3 & 12.1 & 79.8 & 6.8 & 61.6 & 7.1 & 62.6 \\
~ & 4 & ~ & 60.1 & ~ & 90.2 & ~ & 75.9 & ~ & 77.6 \\
~ & 5 & ~ & 49.4 & ~ & 86.1 & ~ & 68.2 & ~ & 69.9 \\
~ & 6 & ~ & 62.9 & ~ & 92.4 & ~ & 79.6 & ~ & 80.7 \\
~ & 7 & ~ & 98.4 & ~ & 100.0 & ~ & 99.7 & ~ & 99.8 \\
\midrule
3 & 5 & 3.0 & 32.2 & 9.1 & 78.9 & 5.7 & 57.9 & 5.8 & 58.3 \\
~ & 6 & ~ & 47.5 & ~ & 89.4 & ~ & 73.2 & ~ & 73.4 \\
~ & 7 & ~ & 91.6 & ~ & 99.8 & ~ & 97.5 & ~ & 97.7 \\
\midrule
4 & 5 & 3.6 & 23.3 & 7.7 & 52.9 & 5.5 & 38.2 & 7.9 & 52.2 \\
~ & 6 & ~ & 39.5 & ~ & 72.6 & ~ & 56.5 & ~ & 70.5 \\
~ & 7 & ~ & 85.8 & ~ & 99.0 & ~ & 90.6 & ~ & 93.1 \\
\midrule
5 & 7 & 2.6 & 72.8 & 4.7 & 93.4 & 4.0 & 90.6 & 4.2 & 90.7 \\
\midrule
6 & 7 & 3.9 & 67.8 & 8.5 & 90.5 & 6.4 & 84.6 & 6.7 & 85.1 \\
\bottomrule
\end{tabular}
}\end{table}
\begin{table}[h!]
\tiny
\centering
\caption{Weighted Conformal Methods: FPR and Power for Different Calibration Sizes and AI-Usage Guidelines. Model: \texttt{Phi-4-mini-instruct}, Method: \texttt{Green/Red List}, NNES calibration
size $m$: 5.}
\label{tab:phi_maryland_n5_ETS_conformal_weighted}
\resizebox{\textwidth}{!}{%
\begin{tabular}{c c | c c | c c | c c | c c }
\toprule
Null & Alt. & \multicolumn{2}{c|}{In Dist. Only} & \multicolumn{2}{c|}{Comb. Unweighted} & \multicolumn{4}{c}{Comb. Weighted}\\
& & & & & & \multicolumn{2}{c|}{Quantile} & \multicolumn{2}{c}{Mean} \\
& & FPR & Power & FPR & Power & FPR & Power & FPR & Power \\
\midrule
1 & 2 & 0.0 & 0.0 & 8.8 & 54.9 & 4.9 & 40.1 & 6.9 & 49.0 \\
~ & 3 & ~ & 0.0 & ~ & 63.9 & ~ & 49.0 & ~ & 58.1 \\
~ & 4 & ~ & 0.0 & ~ & 72.5 & ~ & 59.3 & ~ & 67.1 \\
~ & 5 & ~ & 0.0 & ~ & 77.0 & ~ & 63.5 & ~ & 71.9 \\
~ & 6 & ~ & 0.0 & ~ & 84.4 & ~ & 72.7 & ~ & 80.2 \\
~ & 7 & ~ & 0.0 & ~ & 99.9 & ~ & 99.5 & ~ & 99.8 \\
\midrule
2 & 3 & 0.0 & 0.0 & 10.4 & 62.0 & 3.9 & 32.4 & 7.2 & 48.9 \\
~ & 4 & ~ & 0.0 & ~ & 76.4 & ~ & 47.1 & ~ & 66.2 \\
~ & 5 & ~ & 0.0 & ~ & 69.7 & ~ & 37.7 & ~ & 56.9 \\
~ & 6 & ~ & 0.0 & ~ & 79.6 & ~ & 48.3 & ~ & 68.2 \\
~ & 7 & ~ & 0.0 & ~ & 99.3 & ~ & 90.7 & ~ & 97.7 \\
\midrule
3 & 4 & 0.0 & 0.0 & 8.8 & 75.9 & 4.7 & 52.8 & 7.1 & 69.3 \\
~ & 5 & ~ & 0.0 & ~ & 60.4 & ~ & 37.7 & ~ & 52.5 \\
~ & 6 & ~ & 0.0 & ~ & 74.7 & ~ & 52.0 & ~ & 68.0 \\
~ & 7 & ~ & 0.0 & ~ & 97.8 & ~ & 89.1 & ~ & 96.2 \\
\midrule
4 & 5 & 0.0 & 0.0 & 5.7 & 26.7 & 3.5 & 15.7 & 5.0 & 23.5 \\
~ & 6 & ~ & 0.0 & ~ & 43.1 & ~ & 28.1 & ~ & 38.9 \\
~ & 7 & ~ & 0.0 & ~ & 86.5 & ~ & 67.1 & ~ & 83.2 \\
\midrule
5 & 7 & 0.0 & 0.0 & 5.8 & 79.9 & 3.9 & 66.8 & 4.7 & 75.3 \\
\midrule
6 & 7 & 0.0 & 0.0 & 7.0 & 70.1 & 3.6 & 46.7 & 5.3 & 61.2 \\
\bottomrule
\end{tabular}
}\end{table}
\begin{table}[h!]
\tiny
\centering
\caption{Weighted Conformal Methods: FPR and Power for Different Calibration Sizes and AI-Usage Guidelines. Model: \texttt{Phi-4-mini-instruct}, Method: \texttt{Green/Red List}, NNES calibration
size $m$: 15.}
\label{tab:phi_maryland_n15_ETS_conformal_weighted}
\resizebox{\textwidth}{!}{%
\begin{tabular}{c c | c c | c c | c c | c c }
\toprule
Null & Alt. & \multicolumn{2}{c|}{In Dist. Only} & \multicolumn{2}{c|}{Comb. Unweighted} & \multicolumn{4}{c}{Comb. Weighted}\\
& & & & & & \multicolumn{2}{c|}{Quantile} & \multicolumn{2}{c}{Mean} \\
& & FPR & Power & FPR & Power & FPR & Power & FPR & Power \\
\midrule
1 & 2 & 0.0 & 0.0 & 8.8 & 54.9 & 6.1 & 45.6 & 5.8 & 44.6 \\
~ & 3 & ~ & 0.0 & ~ & 63.8 & ~ & 54.5 & ~ & 53.6 \\
~ & 4 & ~ & 0.0 & ~ & 72.5 & ~ & 63.8 & ~ & 63.0 \\
~ & 5 & ~ & 0.0 & ~ & 77.0 & ~ & 68.6 & ~ & 67.6 \\
~ & 6 & ~ & 0.0 & ~ & 84.4 & ~ & 77.1 & ~ & 76.5 \\
~ & 7 & ~ & 0.0 & ~ & 99.9 & ~ & 99.0 & ~ & 99.0 \\
\midrule
2 & 3 & 0.0 & 0.0 & 10.4 & 62.0 & 6.5 & 46.6 & 6.3 & 46.7 \\
~ & 4 & ~ & 0.0 & ~ & 76.3 & ~ & 63.8 & ~ & 63.9 \\
~ & 5 & ~ & 0.0 & ~ & 69.6 & ~ & 53.9 & ~ & 53.5 \\
~ & 6 & ~ & 0.0 & ~ & 79.5 & ~ & 64.9 & ~ & 64.9 \\
~ & 7 & ~ & 0.0 & ~ & 99.3 & ~ & 96.9 & ~ & 97.0 \\
\midrule
3 & 4 & 0.0 & 0.0 & 9.0 & 75.8 & 6.3 & 61.5 & 6.7 & 62.2 \\
~ & 5 & ~ & 0.0 & ~ & 61.3 & ~ & 46.0 & ~ & 47.3 \\
~ & 6 & ~ & 0.0 & ~ & 75.4 & ~ & 60.8 & ~ & 62.1 \\
~ & 7 & ~ & 0.0 & ~ & 97.9 & ~ & 93.1 & ~ & 93.5 \\
\midrule
4 & 5 & 0.0 & 0.0 & 5.8 & 26.9 & 5.0 & 22.5 & 5.6 & 25.5 \\
~ & 6 & ~ & 0.0 & ~ & 43.3 & ~ & 38.2 & ~ & 41.1 \\
~ & 7 & ~ & 0.0 & ~ & 86.7 & ~ & 82.6 & ~ & 83.6 \\
\midrule
5 & 7 & 0.0 & 0.0 & 5.9 & 80.2 & 4.9 & 75.6 & 4.9 & 74.8 \\
\midrule
6 & 7 & 0.0 & 0.0 & 7.0 & 70.1 & 5.3 & 61.3 & 5.2 & 59.2 \\
\bottomrule
\end{tabular}
}\end{table}
\begin{table}[h!]
\tiny
\centering
\caption{Weighted Conformal Methods: FPR and Power for Different Calibration Sizes and AI-Usage Guidelines. Model: \texttt{Phi-4-mini-instruct}, Method: \texttt{Green/Red List}, NNES calibration
size $m$: 30.}
\label{tab:phi_maryland_n30_ETS_conformal_weighted}
\resizebox{\textwidth}{!}{%
\begin{tabular}{c c | c c | c c | c c | c c }
\toprule
Null & Alt. & \multicolumn{2}{c|}{In Dist. Only} & \multicolumn{2}{c|}{Comb. Unweighted} & \multicolumn{4}{c}{Comb. Weighted}\\
& & & & & & \multicolumn{2}{c|}{Quantile} & \multicolumn{2}{c}{Mean} \\
& & FPR & Power & FPR & Power & FPR & Power & FPR & Power \\
\midrule
1 & 2 & 3.0 & 31.9 & 8.5 & 54.4 & 6.2 & 47.5 & 5.4 & 44.0 \\
~ & 3 & ~ & 39.7 & ~ & 63.1 & ~ & 56.4 & ~ & 52.8 \\
~ & 4 & ~ & 50.8 & ~ & 72.0 & ~ & 65.3 & ~ & 61.9 \\
~ & 5 & ~ & 53.9 & ~ & 76.5 & ~ & 70.3 & ~ & 66.6 \\
~ & 6 & ~ & 64.1 & ~ & 83.9 & ~ & 78.6 & ~ & 75.3 \\
~ & 7 & ~ & 97.1 & ~ & 99.9 & ~ & 99.2 & ~ & 98.5 \\
\midrule
2 & 3 & 3.1 & 29.2 & 9.9 & 60.6 & 6.1 & 44.9 & 5.8 & 43.8 \\
~ & 4 & ~ & 45.2 & ~ & 75.3 & ~ & 61.6 & ~ & 60.6 \\
~ & 5 & ~ & 34.9 & ~ & 68.1 & ~ & 51.9 & ~ & 50.8 \\
~ & 6 & ~ & 45.8 & ~ & 78.1 & ~ & 63.3 & ~ & 61.8 \\
~ & 7 & ~ & 86.9 & ~ & 99.2 & ~ & 95.4 & ~ & 95.0 \\
\midrule
3 & 4 & 3.3 & 39.9 & 8.4 & 74.6 & 6.2 & 63.9 & 5.9 & 61.9 \\
~ & 5 & ~ & 26.4 & ~ & 58.9 & ~ & 47.4 & ~ & 44.9 \\
~ & 6 & ~ & 38.8 & ~ & 73.6 & ~ & 62.3 & ~ & 60.3 \\
~ & 7 & ~ & 78.3 & ~ & 97.6 & ~ & 94.5 & ~ & 92.7 \\
\midrule
4 & 5 & 3.7 & 16.6 & 5.5 & 26.4 & 5.3 & 25.1 & 5.4 & 25.1 \\
~ & 6 & ~ & 28.2 & ~ & 42.3 & ~ & 41.2 & ~ & 40.6 \\
~ & 7 & ~ & 66.7 & ~ & 86.1 & ~ & 85.0 & ~ & 83.9 \\
\midrule
5 & 7 & 2.8 & 56.1 & 5.6 & 79.2 & 5.0 & 76.7 & 4.8 & 75.6 \\
\midrule
6 & 7 & 3.1 & 43.9 & 6.8 & 69.4 & 5.4 & 62.2 & 5.3 & 60.8 \\
\bottomrule
\end{tabular}
}\end{table}
\begin{table}[h!]
\tiny
\centering
\caption{Weighted Conformal Methods: FPR and Power for Different Calibration Sizes and AI-Usage Guidelines. Model: \texttt{Qwen2.5-7B-Instruct}, Method: \texttt{Gumbel-Max}, NNES calibration
size $m$: 5.}
\label{tab:qwen_openai_n5_ETS_conformal_weighted}
\resizebox{\textwidth}{!}{%
\begin{tabular}{c c | c c | c c | c c | c c }
\toprule
Null & Alt. & \multicolumn{2}{c|}{In Dist. Only} & \multicolumn{2}{c|}{Comb. Unweighted} & \multicolumn{4}{c}{Comb. Weighted}\\
& & & & & & \multicolumn{2}{c|}{Quantile} & \multicolumn{2}{c}{Mean} \\
& & FPR & Power & FPR & Power & FPR & Power & FPR & Power \\
\midrule
1 & 2 & 0.0 & 0.0 & 11.1 & 60.2 & 5.2 & 38.7 & 8.1 & 51.6 \\
~ & 3 & ~ & 0.0 & ~ & 86.2 & ~ & 67.6 & ~ & 80.4 \\
~ & 4 & ~ & 0.0 & ~ & 92.0 & ~ & 77.4 & ~ & 87.8 \\
~ & 5 & ~ & 0.0 & ~ & 92.7 & ~ & 77.0 & ~ & 88.1 \\
~ & 6 & ~ & 0.0 & ~ & 97.1 & ~ & 87.0 & ~ & 94.4 \\
~ & 7 & ~ & 0.0 & ~ & 98.3 & ~ & 94.3 & ~ & 97.2 \\
\midrule
2 & 3 & 0.0 & 0.0 & 8.5 & 69.0 & 4.3 & 47.2 & 6.5 & 61.1 \\
~ & 4 & ~ & 0.0 & ~ & 80.8 & ~ & 60.0 & ~ & 74.5 \\
~ & 5 & ~ & 0.0 & ~ & 74.4 & ~ & 51.9 & ~ & 66.9 \\
~ & 6 & ~ & 0.0 & ~ & 87.4 & ~ & 67.8 & ~ & 81.7 \\
~ & 7 & ~ & 0.0 & ~ & 94.1 & ~ & 81.8 & ~ & 91.3 \\
\midrule
3 & 5 & 0.0 & 0.0 & 4.6 & 27.2 & 2.0 & 14.6 & 3.4 & 21.5 \\
~ & 6 & ~ & 0.0 & ~ & 42.7 & ~ & 26.3 & ~ & 36.1 \\
~ & 7 & ~ & 0.0 & ~ & 57.3 & ~ & 37.7 & ~ & 50.0 \\
\midrule
4 & 5 & 0.0 & 0.0 & 3.8 & 10.9 & 2.7 & 7.6 & 3.2 & 9.2 \\
~ & 6 & ~ & 0.0 & ~ & 22.4 & ~ & 15.5 & ~ & 19.4 \\
~ & 7 & ~ & 0.0 & ~ & 32.3 & ~ & 23.8 & ~ & 27.9 \\
\midrule
5 & 7 & 0.0 & 0.0 & 2.2 & 25.9 & 1.3 & 17.6 & 2.0 & 24.4 \\
\midrule
6 & 7 & 0.0 & 0.0 & 2.9 & 14.1 & 1.8 & 8.4 & 2.2 & 10.5 \\
\bottomrule
\end{tabular}
}\end{table}
\begin{table}[h!]
\tiny
\centering
\caption{Weighted Conformal Methods: FPR and Power for Different Calibration Sizes and AI-Usage Guidelines. Model: \texttt{Qwen2.5-7B-Instruct}, Method: \texttt{Gumbel-Max}, NNES calibration
size $m$: 15.}
\label{tab:qwen_openai_n15_ETS_conformal_weighted}
\resizebox{\textwidth}{!}{%
\begin{tabular}{c c | c c | c c | c c | c c }
\toprule
Null & Alt. & \multicolumn{2}{c|}{In Dist. Only} & \multicolumn{2}{c|}{Comb. Unweighted} & \multicolumn{4}{c}{Comb. Weighted}\\
& & & & & & \multicolumn{2}{c|}{Quantile} & \multicolumn{2}{c}{Mean} \\
& & FPR & Power & FPR & Power & FPR & Power & FPR & Power \\
\midrule
1 & 2 & 0.0 & 0.0 & 11.0 & 60.1 & 7.4 & 48.0 & 6.9 & 46.0 \\
~ & 3 & ~ & 0.0 & ~ & 86.1 & ~ & 76.0 & ~ & 74.6 \\
~ & 4 & ~ & 0.0 & ~ & 92.0 & ~ & 83.6 & ~ & 82.5 \\
~ & 5 & ~ & 0.0 & ~ & 92.6 & ~ & 83.6 & ~ & 82.4 \\
~ & 6 & ~ & 0.0 & ~ & 97.1 & ~ & 91.2 & ~ & 90.4 \\
~ & 7 & ~ & 0.0 & ~ & 98.3 & ~ & 95.4 & ~ & 95.1 \\
\midrule
2 & 3 & 0.0 & 0.0 & 8.5 & 69.0 & 5.7 & 56.1 & 6.3 & 57.5 \\
~ & 4 & ~ & 0.0 & ~ & 80.8 & ~ & 69.2 & ~ & 70.5 \\
~ & 5 & ~ & 0.0 & ~ & 74.3 & ~ & 61.1 & ~ & 62.4 \\
~ & 6 & ~ & 0.0 & ~ & 87.3 & ~ & 76.7 & ~ & 77.7 \\
~ & 7 & ~ & 0.0 & ~ & 94.1 & ~ & 87.8 & ~ & 88.8 \\
\midrule
3 & 5 & 0.0 & 0.0 & 4.9 & 28.0 & 4.1 & 24.5 & 5.9 & 29.6 \\
~ & 6 & ~ & 0.0 & ~ & 44.0 & ~ & 39.9 & ~ & 45.6 \\
~ & 7 & ~ & 0.0 & ~ & 58.6 & ~ & 54.0 & ~ & 58.8 \\
\midrule
4 & 5 & 0.0 & 0.0 & 4.0 & 11.3 & 3.8 & 11.0 & 4.7 & 13.0 \\
~ & 6 & ~ & 0.0 & ~ & 23.3 & ~ & 22.6 & ~ & 25.8 \\
~ & 7 & ~ & 0.0 & ~ & 33.3 & ~ & 32.2 & ~ & 35.8 \\
\midrule
5 & 7 & 0.0 & 0.0 & 2.5 & 27.2 & 2.4 & 26.8 & 2.9 & 28.6 \\
\midrule
6 & 7 & 0.0 & 0.0 & 2.9 & 14.2 & 2.7 & 13.1 & 3.6 & 15.3 \\
\bottomrule
\end{tabular}
}\end{table}
\begin{table}[h!]
\tiny
\centering
\caption{Weighted Conformal Methods: FPR and Power for Different Calibration Sizes and AI-Usage Guidelines. Model: \texttt{Qwen2.5-7B-Instruct}, Method: \texttt{Gumbel-Max}, NNES calibration
size $m$: 30.}
\label{tab:qwen_openai_n30_ETS_conformal_weighted}
\resizebox{\textwidth}{!}{%
\begin{tabular}{c c | c c | c c | c c | c c }
\toprule
Null & Alt. & \multicolumn{2}{c|}{In Dist. Only} & \multicolumn{2}{c|}{Comb. Unweighted} & \multicolumn{4}{c}{Comb. Weighted}\\
& & & & & & \multicolumn{2}{c|}{Quantile} & \multicolumn{2}{c}{Mean} \\
& & FPR & Power & FPR & Power & FPR & Power & FPR & Power \\
\midrule
1 & 2 & 2.3 & 27.0 & 10.6 & 59.2 & 5.9 & 44.2 & 5.4 & 42.3 \\
~ & 3 & ~ & 56.0 & ~ & 85.8 & ~ & 74.3 & ~ & 73.0 \\
~ & 4 & ~ & 67.6 & ~ & 91.7 & ~ & 82.8 & ~ & 81.6 \\
~ & 5 & ~ & 66.4 & ~ & 92.4 & ~ & 82.7 & ~ & 81.6 \\
~ & 6 & ~ & 79.7 & ~ & 96.9 & ~ & 91.0 & ~ & 90.2 \\
~ & 7 & ~ & 91.0 & ~ & 98.3 & ~ & 95.7 & ~ & 95.3 \\
\midrule
2 & 3 & 2.6 & 38.2 & 7.6 & 66.7 & 5.4 & 56.5 & 5.8 & 57.7 \\
~ & 4 & ~ & 51.7 & ~ & 79.0 & ~ & 70.3 & ~ & 71.1 \\
~ & 5 & ~ & 43.1 & ~ & 72.3 & ~ & 61.9 & ~ & 63.0 \\
~ & 6 & ~ & 60.3 & ~ & 86.1 & ~ & 77.7 & ~ & 78.6 \\
~ & 7 & ~ & 77.8 & ~ & 93.7 & ~ & 89.1 & ~ & 89.6 \\
\midrule
3 & 5 & 4.3 & 22.7 & 4.6 & 26.8 & 4.4 & 26.1 & 5.3 & 28.4 \\
~ & 6 & ~ & 36.6 & ~ & 42.3 & ~ & 41.5 & ~ & 44.5 \\
~ & 7 & ~ & 48.2 & ~ & 57.0 & ~ & 56.1 & ~ & 58.3 \\
\midrule
4 & 5 & 2.5 & 7.2 & 3.8 & 10.9 & 3.7 & 10.5 & 4.6 & 12.7 \\
~ & 6 & ~ & 15.7 & ~ & 22.4 & ~ & 21.9 & ~ & 25.4 \\
~ & 7 & ~ & 21.6 & ~ & 32.2 & ~ & 31.4 & ~ & 35.5 \\
\midrule
5 & 7 & 3.2 & 28.1 & 2.4 & 26.8 & 2.3 & 26.5 & 2.4 & 26.5 \\
\midrule
6 & 7 & 3.0 & 12.3 & 2.9 & 14.1 & 2.8 & 13.7 & 3.2 & 14.7 \\
\bottomrule
\end{tabular}
}\end{table}
\begin{table}[h!]
\tiny
\centering
\caption{Weighted Conformal Methods: FPR and Power for Different Calibration Sizes and AI-Usage Guidelines. Model: \texttt{Qwen2.5-7B-Instruct}, Method: \texttt{Green/Red List}, NNES calibration
size $m$: 5.}
\label{tab:qwen_maryland_n5_ETS_conformal_weighted}
\resizebox{\textwidth}{!}{%
\begin{tabular}{c c | c c | c c | c c | c c }
\toprule
Null & Alt. & \multicolumn{2}{c|}{In Dist. Only} & \multicolumn{2}{c|}{Comb. Unweighted} & \multicolumn{4}{c}{Comb. Weighted}\\
& & & & & & \multicolumn{2}{c|}{Quantile} & \multicolumn{2}{c}{Mean} \\
& & FPR & Power & FPR & Power & FPR & Power & FPR & Power \\
\midrule
1 & 2 & 0.0 & 0.0 & 7.6 & 41.0 & 4.9 & 31.3 & 6.3 & 37.2 \\
~ & 3 & ~ & 0.0 & ~ & 67.4 & ~ & 57.5 & ~ & 63.6 \\
~ & 4 & ~ & 0.0 & ~ & 77.1 & ~ & 67.9 & ~ & 74.0 \\
~ & 5 & ~ & 0.0 & ~ & 75.4 & ~ & 65.9 & ~ & 72.1 \\
~ & 6 & ~ & 0.0 & ~ & 86.7 & ~ & 79.4 & ~ & 84.2 \\
~ & 7 & ~ & 0.0 & ~ & 97.7 & ~ & 95.2 & ~ & 96.9 \\
\midrule
2 & 3 & 0.0 & 0.0 & 9.0 & 55.4 & 4.7 & 36.8 & 7.4 & 50.2 \\
~ & 4 & ~ & 0.0 & ~ & 65.9 & ~ & 46.2 & ~ & 60.4 \\
~ & 5 & ~ & 0.0 & ~ & 59.4 & ~ & 40.5 & ~ & 54.3 \\
~ & 6 & ~ & 0.0 & ~ & 74.5 & ~ & 54.8 & ~ & 69.6 \\
~ & 7 & ~ & 0.0 & ~ & 92.0 & ~ & 79.0 & ~ & 89.3 \\
\midrule
3 & 5 & 0.0 & 0.0 & 8.1 & 28.3 & 2.8 & 11.6 & 4.6 & 18.0 \\
~ & 6 & ~ & 0.0 & ~ & 39.8 & ~ & 18.2 & ~ & 27.2 \\
~ & 7 & ~ & 0.0 & ~ & 62.8 & ~ & 33.2 & ~ & 47.5 \\
\midrule
4 & 5 & 0.0 & 0.0 & 5.5 & 11.2 & 3.0 & 5.9 & 4.2 & 8.6 \\
~ & 6 & ~ & 0.0 & ~ & 19.4 & ~ & 11.3 & ~ & 15.5 \\
~ & 7 & ~ & 0.0 & ~ & 37.3 & ~ & 22.6 & ~ & 31.0 \\
\midrule
5 & 6 & 0.0 & 0.0 & 3.4 & 14.7 & 2.4 & 10.5 & 3.0 & 13.2 \\
~ & 7 & ~ & 0.0 & ~ & 29.7 & ~ & 22.7 & ~ & 27.2 \\
\midrule
6 & 7 & 0.0 & 0.0 & 4.1 & 20.9 & 2.5 & 14.0 & 2.9 & 16.0 \\
\bottomrule
\end{tabular}
}\end{table}
\begin{table}[h!]
\tiny
\centering
\caption{Weighted Conformal Methods: FPR and Power for Different Calibration Sizes and AI-Usage Guidelines. Model: \texttt{Qwen2.5-7B-Instruct}, Method: \texttt{Green/Red List}, NNES calibration
size $m$: 15.}
\label{tab:qwen_maryland_n15_ETS_conformal_weighted}
\resizebox{\textwidth}{!}{%
\begin{tabular}{c c | c c | c c | c c | c c }
\toprule
Null & Alt. & \multicolumn{2}{c|}{In Dist. Only} & \multicolumn{2}{c|}{Comb. Unweighted} & \multicolumn{4}{c}{Comb. Weighted}\\
& & & & & & \multicolumn{2}{c|}{Quantile} & \multicolumn{2}{c}{Mean} \\
& & FPR & Power & FPR & Power & FPR & Power & FPR & Power \\
\midrule
1 & 2 & 0.0 & 0.0 & 7.6 & 41.1 & 6.1 & 35.5 & 6.0 & 34.7 \\
~ & 3 & ~ & 0.0 & ~ & 67.5 & ~ & 60.5 & ~ & 59.4 \\
~ & 4 & ~ & 0.0 & ~ & 77.3 & ~ & 70.5 & ~ & 69.5 \\
~ & 5 & ~ & 0.0 & ~ & 75.4 & ~ & 68.5 & ~ & 67.6 \\
~ & 6 & ~ & 0.0 & ~ & 86.7 & ~ & 80.3 & ~ & 79.4 \\
~ & 7 & ~ & 0.0 & ~ & 97.7 & ~ & 94.2 & ~ & 93.8 \\
\midrule
2 & 3 & 0.0 & 0.0 & 9.0 & 55.4 & 7.1 & 47.3 & 7.3 & 47.8 \\
~ & 4 & ~ & 0.0 & ~ & 65.8 & ~ & 57.3 & ~ & 57.7 \\
~ & 5 & ~ & 0.0 & ~ & 59.4 & ~ & 51.4 & ~ & 52.0 \\
~ & 6 & ~ & 0.0 & ~ & 74.5 & ~ & 66.2 & ~ & 66.6 \\
~ & 7 & ~ & 0.0 & ~ & 92.0 & ~ & 86.8 & ~ & 86.8 \\
\midrule
3 & 5 & 0.0 & 0.0 & 8.2 & 28.3 & 6.1 & 21.6 & 6.0 & 20.7 \\
~ & 6 & ~ & 0.0 & ~ & 40.1 & ~ & 32.5 & ~ & 31.0 \\
~ & 7 & ~ & 0.0 & ~ & 63.0 & ~ & 52.7 & ~ & 50.6 \\
\midrule
4 & 5 & 0.0 & 0.0 & 5.5 & 11.1 & 4.4 & 8.6 & 4.9 & 9.8 \\
~ & 6 & ~ & 0.0 & ~ & 19.5 & ~ & 16.2 & ~ & 17.2 \\
~ & 7 & ~ & 0.0 & ~ & 37.4 & ~ & 31.9 & ~ & 33.7 \\
\midrule
5 & 6 & 0.0 & 0.0 & 3.5 & 15.2 & 3.4 & 15.0 & 3.8 & 15.4 \\
~ & 7 & ~ & 0.0 & ~ & 30.5 & ~ & 30.1 & ~ & 31.4 \\
\midrule
6 & 7 & 0.0 & 0.0 & 4.1 & 21.1 & 3.4 & 17.9 & 3.8 & 18.8 \\
\bottomrule
\end{tabular}
}\end{table}
\begin{table}[h!]
\tiny
\centering
\caption{Weighted Conformal Methods: FPR and Power for Different Calibration Sizes and AI-Usage Guidelines. Model: \texttt{Qwen2.5-7B-Instruct}, Method: \texttt{Green/Red List}, NNES calibration
size $m$: 30.}
\label{tab:qwen_maryland_n30_ETS_conformal_weighted}
\resizebox{\textwidth}{!}{%
\begin{tabular}{c c | c c | c c | c c | c c }
\toprule
Null & Alt. & \multicolumn{2}{c|}{In Dist. Only} & \multicolumn{2}{c|}{Comb. Unweighted} & \multicolumn{4}{c}{Comb. Weighted}\\
& & & & & & \multicolumn{2}{c|}{Quantile} & \multicolumn{2}{c}{Mean} \\
& & FPR & Power & FPR & Power & FPR & Power & FPR & Power \\
\midrule
1 & 2 & 3.3 & 23.4 & 7.4 & 40.6 & 6.2 & 36.7 & 5.7 & 34.8 \\
~ & 3 & ~ & 47.1 & ~ & 67.0 & ~ & 62.0 & ~ & 59.8 \\
~ & 4 & ~ & 57.5 & ~ & 76.8 & ~ & 72.2 & ~ & 70.0 \\
~ & 5 & ~ & 55.7 & ~ & 74.9 & ~ & 70.1 & ~ & 67.8 \\
~ & 6 & ~ & 69.2 & ~ & 86.4 & ~ & 81.8 & ~ & 79.8 \\
~ & 7 & ~ & 88.7 & ~ & 97.6 & ~ & 94.7 & ~ & 93.8 \\
\midrule
2 & 3 & 3.4 & 30.1 & 8.6 & 54.2 & 6.6 & 46.7 & 6.0 & 44.4 \\
~ & 4 & ~ & 39.4 & ~ & 64.8 & ~ & 57.0 & ~ & 54.6 \\
~ & 5 & ~ & 34.2 & ~ & 58.3 & ~ & 51.0 & ~ & 49.0 \\
~ & 6 & ~ & 47.8 & ~ & 73.7 & ~ & 66.1 & ~ & 64.0 \\
~ & 7 & ~ & 73.1 & ~ & 91.6 & ~ & 86.7 & ~ & 85.3 \\
\midrule
3 & 5 & 3.2 & 12.6 & 7.8 & 27.5 & 5.6 & 21.3 & 6.5 & 23.1 \\
~ & 6 & ~ & 19.1 & ~ & 39.3 & ~ & 31.1 & ~ & 33.6 \\
~ & 7 & ~ & 34.7 & ~ & 62.1 & ~ & 52.2 & ~ & 54.8 \\
\midrule
4 & 5 & 2.8 & 5.4 & 5.4 & 11.0 & 5.0 & 9.7 & 5.5 & 10.4 \\
~ & 6 & ~ & 10.7 & ~ & 19.1 & ~ & 17.9 & ~ & 18.9 \\
~ & 7 & ~ & 22.0 & ~ & 37.0 & ~ & 34.5 & ~ & 36.1 \\
\midrule
5 & 6 & 3.3 & 13.2 & 3.5 & 14.9 & 3.3 & 14.2 & 3.6 & 14.8 \\
~ & 7 & ~ & 27.1 & ~ & 30.0 & ~ & 29.1 & ~ & 30.8 \\
\midrule
6 & 7 & 2.6 & 12.6 & 3.9 & 20.4 & 3.6 & 18.9 & 3.5 & 18.4 \\
\bottomrule
\end{tabular}
}\end{table}

\end{document}